\documentclass[twocolumn,10pt]{article}

\usepackage{arxiv}

\usepackage[utf8]{inputenc} 
\usepackage[T1]{fontenc}    

\usepackage{url}            

\usepackage{amsfonts}       
\usepackage{nicefrac}       
\usepackage{microtype}      
\usepackage{graphicx}
\usepackage[numbers]{natbib}

\usepackage{amssymb}
\usepackage{amsmath}
\usepackage{amsthm}
\usepackage{tabularx}
\usepackage{booktabs} 
\usepackage{float}
\usepackage{makecell}
\usepackage{rotating}
\usepackage{xcolor} 
\usepackage[normalem]{ulem}
\usepackage{siunitx}
\usepackage{mathrsfs}  
\usepackage{mathtools}
\usepackage{graphicx}
\usepackage{array}
\usepackage[linktoc=page,
colorlinks=true,	
linkcolor=blue,
citecolor=blue,
urlcolor=blue]{hyperref}       

\usepackage{doi}

\usepackage{newtxtext}
\usepackage{newtxmath}
\usepackage{upgreek}
\usepackage{enumitem}
\usepackage{silence}
\usepackage{stmaryrd}
\usepackage{siunitx}
\usepackage{upgreek}
\usepackage{algorithm}
\usepackage{algpseudocode}
\usepackage{subcaption}
\usepackage{float} 
\usepackage{ulem} 
\usepackage{amsthm}

\usepackage{chngcntr}

\newcommand{\ve}[1]{\boldsymbol{#1}} 
\newcommand{\te}[1]{\boldsymbol{\mathrm{#1}}} 
\newcommand{\tte}[1]{\mathbb #1} 
\newcommand{\pd}[2]{\frac{\partial #1}{\partial #2}} 
\newcommand{\jump}[1]{\llbracket #1\rrbracket}

\newcommand{\dd}{\mathrm{d}}

\newcommand{\op}[1]{\operatorname{#1}}

\newcommand{\norm}[1]{\lvert #1 \rvert}
\newcommand{\frob}[1]{\left\lVert #1 \right\rVert}
\newcommand{\ho}[1]{\bigl< #1 \bigr>}

\newcommand{\grad}{\nabla}
\newcommand{\Grad}{\nabla_{\!\!X}}
\renewcommand{\div}{\nabla\cdot}

\newcommand{\rot}{\nabla\times}
\newcommand{\Rot}{\nabla_{\!\!X}\times}

\newcommand{\Ptot}{\te{P}^\text{tot}}
\newcommand{\sigtot}{\te{\upsigma}^\text{tot}}

\theoremstyle{remark}
\newtheorem{remark}{Remark}

\title{Generating Accurate Microstructures of Anisotropic Magnetorheological Elastomers}

\author{
	Heinrich T. Roth\\
	Institute of Solid Mechanics\\
	TU Dresden,
	01069 Dresden, Germany \\
    \And
	Lukas Jabs\\
	Institute of Engineering Mathematics \\
	University of Duisburg-Essen,
	45141 Essen, Germany \\
    \And
	Matti Schneider\\
	Institute of Engineering Mathematics \\
	University of Duisburg-Essen,
	45141 Essen, Germany \\
    \And
	Konstantin Gerstenberger\\
	Leibniz Institute of Polymer Research Dresden \\
	01069 Dresden, Germany \\
    \And
	Günter K. Auernhammer\\
	Leibniz Institute of Polymer Research Dresden \\
	01069 Dresden, Germany \\
    \And
    Karl A. Kalina\\
	Institute of Solid Mechanics\\
	TU Dresden,
	01069 Dresden, Germany \\
	\And
	Markus K\"{a}stner$^{*}$\\
	Institute of Solid Mechanics\\
	TU Dresden, 
	01069 Dresden, Germany \\
}

\renewcommand{\shorttitle}{Generating Accurate Microstructures of Anisotropic Magnetorheological Elastomers}

\hypersetup{
	pdftitle={Preprint\_RothEtAl\_2026},
	pdfsubject={},
	pdfauthor={RothEtAl},
	pdfkeywords={}
}

\usepackage[font=small]{caption} 

\begin{document}
\twocolumn[
\begin{@twocolumnfalse}
    \maketitle
    \begin{abstract}
        Anisotropic magnetorheological elastomers are magnetoactive smart materials whose complex magneto-mechanical behavior is strongly influenced by their underlying microstructure. 
        Efficient and accurate multiscale simulations of these materials rely on small unit cells that reproduce the behavior of the real material. 
        In this work, we present a framework for the generation and validation of such two-dimensional representative volume elements (RVEs) from a given microstructure image, assuming circular particles. 
        The cells are generated through an optimization procedure where the two-point correlation function serves as a differentiable descriptor, while a minimum inter-particle distance is enforced as a constraint. 
        To assess the equivalence of the generated cells, we compare their effective magneto-mechanical response with those of the reference microstructure.
        We demonstrate that the generated cells reproduce the reference response with high accuracy, also with substantially fewer particles.
        Consequently, the computational cost of subsequent multiscale simulations is reduced.
    \end{abstract}
    \vspace{0.5cm}
\end{@twocolumnfalse}
]
{
\renewcommand{\thefootnote}{}
\footnotetext{$^{*}$Corresponding author, email:\\ \texttt{markus.kaestner@tu-dresden.de}.}
}

\section{Introduction} 

Magnetorheological elastomers (MREs) are magnetoactive smart materials whose mechanical properties and shape can be adapted by an external magnetic field \cite{Bastola20,Cantera17,Pelteret2020}. 
This field-dependent behavior enables a broad range of applications:
MREs serve as actuators \cite{Moreno-Mateos22,Becker2019}, for example as magnetically actuated lamellae \cite{Kravanja2023} or tubes \cite{Zhang2025} for object transport and as valves \cite{Boese12}, as well as in magnetically driven robots \cite{Kim19,Dewang2025}. 
They are also used in sensors such as acceleration sensors \cite{Volkova17,Becker2018} and magnetometers \cite{Du12}.
Furthermore, their adjustable stiffness is utilized in adaptive dampers \cite{Kim06,Li13,Lin06,Rasooli2020}.

Structurally, an MRE is a composite material consisting of a mechanically soft elastomeric matrix filled with stiff magnetic particles.
The particles are called magnetically hard in case they exhibit a magnetic hysteresis and soft otherwise.
In most applications, carbonyl iron powder (CIP) with a particle diameter of approximately \SI{5}{\micro\meter} is used as the soft-magnetic phase \cite{Bastola20}.
For research purposes, larger nickel particles with a more regular shape and diameters of \SI{50}{\micro\meter} and above are also used \cite{Schmidt2021_collision,Schmidt2021}.

The particle arrangement within the matrix determines the resulting material behavior. 
If no magnetic field is applied during crosslinking of the elastomer, then the particles remain randomly distributed, yielding an isotropic MRE. 
Conversely, if a magnetic field is applied while the matrix is curing, then the particles align into chain-like structures, resulting in an anisotropic, structured MRE. 
The effective magneto-mechanical behavior depends strongly on this particle arrangement and is therefore of particular interest \cite{Yao2019,Borin2026}.

The effective behavior of complex heterogeneous materials is commonly obtained by solving the governing field equations within a continuum formulation.
For a sample or part made from an MRE, however, a direct finite element (FE) simulation resolving the entire microstructure that may contain millions of particles entails a prohibitive computational effort.
The effort may be reduced by using the theory of homogenization, which makes it possible to predict the effective response of a composite from knowing the microstructure and the behavior of the phases.
In a coupled multiscale scheme \cite{Schroeder13,Keip16,Zabihyan18} only a small portion of the microstructure, a representative volume element (RVE) \cite{Hill63}, is simulated and the local fields are homogenized \cite{Chatzigeorgiou14} to obtain the effective behavior for the macroscopic simulation. 
A decoupled multiscale scheme \cite{Gebhart22_2,Kalina20-2,Roth2026} may be even faster: a macroscopic material model is calibrated on homogenized RVE data and then used in the macroscopic simulations. 
In both the coupled and decoupled cases, the volume element should be as small as possible while still being large enough to be representative of the microstructure. 
The use of a periodic RVE together with periodic boundary conditions is additionally favorable for achieving representativity~\cite{Sab2005,Ravichandran2026}.

The field of microstructure generation aims to generate unit cells that exhibit prescribed morphological features, which may be quantified by a descriptor~\cite{Bargmann2018,Bostanabad2018}.
The generation may be formulated as an optimization problem, in which a unit cell is obtained by minimizing the error between its descriptor and the descriptor of a reference microstructure.
A commonly used descriptor is the two-point correlation (TPC), which characterizes the spatial neighborhood relations between the phases of a microstructure~\cite{Yeong1998,Torquato2002}.
Since it does not presuppose a particular microstructure geometry, the TPC is applicable to a broad range of heterogeneous materials.
In certain settings, however, descriptors tailored to the specific microstructure at hand have proven effective for generation \cite{Schneider2017,Mehta2022}.
In many cases, the microstructure is simply described by pixels or voxels.
Since the number of parameters is given by the number of pixels or voxels, the generation of larger cells is computationally expensive, even with gradient-based optimization approaches \cite{Seibert2021,SeibertDiss}.
Furthermore, it is not guaranteed that the resulting cell satisfies certain conditions on the shape, size, and minimum distance between particles, which might complicate subsequent microscale FE simulations.

To avoid these drawbacks, we take the efficient and differentiable formulation of the TPC for ellipsoidal particles, introduced by Seibert~et~al.~\cite{Seibert24}, as our starting point.
Relying on simple geometric shapes, this formulation allows us to constrain the range of particle sizes and distances, while at the same time significantly reducing the number of optimization parameters.
This paper presents a method to generate two-dimensional statistically equivalent unit cells from a given two-dimensional microstructure image using the TPC as the given descriptor.
To validate the statistical equivalence between generated and reference cells, we compare their effective magneto-mechanical behavior.

\begin{remark}
    Although real microstructures are three-dimen\-sional and their particles approximately spherical, a fully three-dimensional treatment is computationally demanding. The present work therefore adopts a two-dimensional formulation as a simplified model system, in which particles are represented as circles, corresponding to infinitely long cylinders in three dimensions. This reduces the computational cost and enables faster evaluation while retaining the essential methodological aspects. Because geometric constraints and particle interactions differ in three dimensions, the quantitative results cannot be expected to transfer directly. The methodology itself, however, should be transferable to three-dimensional microstructures with appropriate modifications.
\end{remark}

This work is outlined as follows. Sec.~\ref{sec:Microstructure_generation} introduces the two-point correlation function as a descriptor and describes the microstructure generation procedure. 
The underlying equations for the magneto-mechanical microscale simulations and the homogenization are presented in Sec.~\ref{sec:Finite_strain_magneto-hyperelasticity}. 
In Sec.~\ref{sec:Numerical_examples}, a microstructure image is employed to generate unit cells, and their effective behavior is subsequently compared. 
Finally, Sec.~\ref{sec:Summary_and_outlook} summarizes the results and outlines future steps.

\paragraph{Notation.}
Throughout this work, indices carry four different meanings depending on the context.

\textit{Tensor indices} are denoted by Greek letters and label the components of a tensor
with respect to a chosen Cartesian basis, e.g., $\ve{a} = \sum_{\alpha=1}^{n_\text{dim}} a_\alpha\, \ve{e}_\alpha$ and $\te{A} = \sum_{\alpha,\beta=1}^{n_\text{dim}} A_{\alpha\beta}\, \ve{e}_\alpha \otimes \ve{e}_\beta$, where $n_\text{dim} \in \{2,3\}$ is the spatial dimension. 
First-order tensors are written in italic bold ($\ve{a}$), second-order tensors in upright bold ($\te{A}$), and fourth-order tensors in blackboard bold ($\tte{A}$). The Einstein summation convention is \textit{not} employed; all summations are stated explicitly.
A non-tensorial quantity associated with the spatial direction $\ve{e}_\alpha$, such as the edge length of a cube aligned with the coordinate axes, is denoted by $a_{(\alpha)}$.

The single contraction is defined as $\ve{a} \cdot \ve{b} = \sum_{\alpha} a_\alpha b_\alpha$ and $\te{A} \cdot \te{B} = \sum_{\alpha,\beta,\gamma} A_{\alpha\beta} B_{\beta\gamma}\, \ve{e}_\alpha \otimes \ve{e}_\gamma$, the double contraction as $\te{A} : \te{B} = \sum_{\alpha,\beta} A_{\alpha\beta} B_{\beta\alpha}$, and the cross product as $\ve{a} \times \ve{b} = \sum_{\alpha,\beta,\gamma} e_{\alpha\beta\gamma}\, a_\alpha b_\beta\, \ve{e}_\gamma$ with the Levi--Civita symbol $e_{\alpha\beta\gamma}$.
The dyadic product is denoted by the symbol $\otimes$.

A second-order tensor admits the additive decompositions $\te{A} = \op{dev}(\te{A}) + \op{sph}(\te{A}) = \op{sym}(\te{A}) + \op{skw}(\te{A})$ into its deviatoric part $\op{dev}(\te{A}) = \te{A} - \tfrac{1}{3}\op{tr}(\te{A})\,\te{1}$ and its volumetric part $\op{sph}(\te{A}) = \tfrac{1}{3}\op{tr}(\te{A})\,\te{1}$, as well as into its symmetric part $\op{sym}(\te{A}) = \tfrac{1}{2}\!\left(\te{A} + \te{A}^{T}\right)$ and skew-symmetric part $\op{skw}(\te{A}) = \tfrac{1}{2}\!\left(\te{A} - \te{A}^{T}\right)$. 
The trace and determinant of a second-order tensor are denoted by $\op{tr}(\te{A}) = \sum_{\alpha} A_{\alpha\alpha}$ and $\op{det}(\te{A})$, respectively.

Using the nabla operator $\nabla$, the gradient $\nabla_{\!X}\, \te{A} = \sum_{\alpha} \ve{e}_\alpha \otimes \pd{\te{A}}{X_\alpha}$, divergence $\nabla_{\!X} \cdot \te{A} = \sum_{\alpha} \ve{e}_\alpha \cdot \pd{\te{A}}{X_\alpha}$, and curl $\nabla_{\!X} \times \te{A} = \sum_{\alpha} \ve{e}_\alpha \times \pd{\te{A}}{X_\alpha}$ are defined with respect to the reference coordinates $X_\alpha$. 
The corresponding operators $\nabla \te{A}$, $\nabla \cdot \te{A}$, and $\nabla \times \te{A}$ are defined analogously with respect to the current coordinates $x_\alpha$. 
Finally, the norm $\frob{\ve{a}} = \sqrt{\sum_{\alpha} a_\alpha a_\alpha}$ and $\frob{\te{A}} = \sqrt{\sum_{\alpha,\beta} A_{\alpha\beta} A_{\alpha\beta}}$
applies to both tensors and arrays.

\textit{Array indices} are denoted by $l, q$, e.g., \\
$\underline{a} = \begin{bmatrix} a_1, \dots, a_l, \dots, a_n \end{bmatrix}$ 
with
$l \in \{1, \dots, n\}$
refering to elements of arrays of size $n \in \mathbb{N}$, which do not possess the transformation properties of a tensor. 

\textit{Grid indices in real space} are denoted by $u,v$ and enumerate the values of a field at discrete spatial grid points, e.g., $a_u = a\!\left( \ve{r}_u \right)$ is the scalar value of the field $a$ at the point $\ve{r}_u$.

\textit{Grid indices in Fourier space} are denoted by $j,k,s,t$ and enumerate quantities at discrete frequencies in Fourier space, e.g., $\hat{a}_j = \hat{a}\!\left( \boldsymbol{\omega}_j \right)$ is the Fourier coefficient for the wave vector $\boldsymbol{\omega}_j$.

Although these conventions are given here for a single index, they extend to several
indices by adding further underlines. 
Arrays of tensors are likewise admissible, e.g., $\underline{\ve{a}} = \begin{bmatrix} \ve{a}_1, \dots, \ve{a}_l, \dots, \ve{a}_n \end{bmatrix}$ with the indices $a_{l\alpha}$.

\section{Microstructure generation}
\label{sec:Microstructure_generation}

In this section, we formulate the descriptor-based microstructure generation procedure. First, we introduce an analytical and differentiable formulation of the TPC function as a descriptor for a periodic two-dimensional unit cell composed of non-overlapping circular inclusions.
We then outline the optimization procedure employed for microstructure generation. 
In App.~\ref{app:Microstructure_generation_with_Spectral_Density_Function}, we investigate the use of the spectral density function (SDF) as an alternative descriptor for microstructure generation.

\subsection{Spectral density function and two-point correlation}

In the following, we use the TPC function to describe the microstructure \cite{Torquato1982,Yeong1998,Torquato2002}. This descriptor is commonly applied to microstructures discretized by pixels or voxels.
Here, however, we apply it to a geometric description of the microstructure, more precisely, a collection of circles, representing the inclusions.
To this end, we simplify the equations derived by Seibert~et~al.~\cite{Seibert24} for ellipsoidal inclusions.

We consider the closure of the periodic domain $\Omega^\text{cell} = [0,L_1]\times[0,L_2] \subset \mathbb{R}^2$, with the spatial lengths $L_1,L_2\in\mathbb{R}_{>0}$, area $A^\text{cell} = L_1 L_2$, and the disjoint measurable subsets of the particle phase $\Omega^\text{incl}$ and matrix phase $\Omega^\text{matrix}$, which define the decomposition $\Omega^\text{cell} = \Omega^\text{incl} \cup \Omega^\text{matrix}$.
We use the indicator function
\begin{equation}
    \mathcal{I} : \Omega^{\mathrm{cell}} \to \left\{0, 1\right\}, \quad
    \ve{x} \mapsto \mathcal{I}(\ve{x}) =
    \begin{cases}
        1, & \ve{x} \in \Omega^{\mathrm{incl}}, \\
        0, & \text{otherwise},
    \end{cases}
\end{equation}
to differentiate between inclusions and the matrix. 
The area fraction of the inclusions is determined via $\phi = \frac{1}{A^\text{cell}} \int_{\Omega^\text{cell}} \mathcal{I}(\ve{x}) \,\dd\ve{x}$.
Since the cell is extended periodically to $\mathbb{R}^2$, the extension of the indicator function is periodic as well.
The two-point autocorrelation of the inclusions
\begin{equation}
    S_2(\ve{r}) = \frac{1}{A^\text{cell}} \int_{\Omega^\text{cell}} \mathcal{I}(\ve{x}) \cdot \mathcal{I}(\ve{x} + \ve{r}) \, \dd \ve{x}
\end{equation}
represents the probability of finding an inclusion at both the start and the end of a randomly placed vector $\ve{r}$.
Since both the indicator function and the TPC are periodic and integrable over the unit cell, they may be represented as the Fourier series
\begin{subequations}
    \begin{align}
        \mathcal{I}(\ve{x}) &= \sum_{j=-\infty}^{\infty} \sum_{k=-\infty}^{\infty} \hat{\mathcal{I}}_{jk} \,\mathrm{e}^{\mathrm{i} \boldsymbol{\omega}_{jk} \cdot \ve{x}} \quad\text{and}\\
        S_2(\ve{r}) &= \sum_{j=-\infty}^{\infty} \sum_{k=-\infty}^{\infty} \hat{S}_{2_{jk}} \,\mathrm{e}^{\mathrm{i} \boldsymbol{\omega}_{jk} \cdot \ve{r}}
    \end{align}
\end{subequations}
with the Fourier coefficients
\begin{subequations}
    \begin{align}
        \hat{\mathcal{I}}_{jk} &= \frac{1}{A^\text{cell}} \int_{\Omega^\text{cell}} \mathcal{I}(\ve{x}) \,\mathrm{e}^{-\mathrm{i} \boldsymbol{\omega}_{jk} \cdot \ve{x}} \, \dd \ve{x} \quad\text{and}\\
        \hat{S}_{2_{jk}} &= \frac{1}{A^\text{cell}} \int_{\Omega^\text{cell}} S_2(\ve{r}) \,\mathrm{e}^{-\mathrm{i} \boldsymbol{\omega}_{jk} \cdot \ve{r}} \, \dd \ve{r}
    \end{align}
\end{subequations}
and the frequency vector 
\begin{equation}
    \boldsymbol{\omega}_{jk} = 
	\boldsymbol{\omega}(j,k) = 
	\frac{2\pi j}{L_1} \ve{e}_1 + \frac{2\pi k}{L_2} \ve{e}_2
\end{equation}
with $j,k \in \mathbb{Z}$.
It follows that $\hat{\mathcal{I}}_{00} = \phi$ holds.
By periodicity, values on all of $\mathbb{R}^2$ are determined by those on $\ve{r} \in \left[-\frac{L_1}{2}, \frac{L_1}{2}\right] \times \left[-\frac{L_2}{2}, \frac{L_2}{2}\right]$.
The Fourier coefficients 
\begin{equation}
	\hat{\mathcal{I}}_{jk}\left(\underline{R}, \underline{\ve{p}}\right) 
	= \sum_{l=1}^{N^\text{incl}} \hat{\mathcal{I}}^\text{incl}_{jk}\left(R_l, \ve{p}_l\right)
\end{equation}
of the entire cell are computed as the sum of the Fourier coefficients of the individual inclusions
\begin{equation}
	\hat{\mathcal{I}}_{jk}^\text{incl}\left(R_l, \ve{p}_l\right) =
	2 \,\mathrm{e}^{- \mathrm{i} \boldsymbol{\omega}_{jk} \cdot \ve{p}_l} \frac{A^\text{incl}_l}{A^\text{cell}} \frac{J_1 \! \left(R_l \lVert\boldsymbol{\omega}_{jk}\rVert\right)}{R_l \lVert\boldsymbol{\omega}_{jk}\rVert}, 
\end{equation}
where $N^\text{incl}$ is the number of particle centers in the cell, $J_1$ denotes the Bessel function of the first kind, and $R_l$, $\ve{p}_l$, and $A^\text{incl}_l = \pi R_l^2$ are the radius, center point, and area of the inclusion with index $l$, respectively.
To compute the zero mode, we use the limit
\begin{equation}
	\lim_{\boldsymbol{\omega} \to \ve{0}} 
	\left(\frac{J_1(R_l \lVert\boldsymbol{\omega}\rVert)}{R_l \lVert\boldsymbol{\omega}\rVert}\right) = \frac{1}{2} ,
\end{equation}
which yields
\begin{equation}
	\hat{\mathcal{I}}_{00}^\text{incl}\left(R_l\right) = \frac{A^\text{incl}_l}{A^\text{cell}} .
\end{equation}
The Fourier coefficients of the TPC, $\hat{S}_{2_{jk}} = \hat{\mathcal{I}}_{jk} \hat{\mathcal{I}}_{jk}^{*} = \norm{\hat{\mathcal{I}}_{jk}}^2$,
are the squared absolute values of the Fourier coefficients of the indicator function with $(\circ)^{*}$ as the complex conjugate, see Seibert~et~al.~\cite{Seibert24} for details.
The coefficients $\norm{\hat{\mathcal{I}}_{jk}}^2$ are also known as the SDF.
The derived expressions for the indicator function and the TPC are only valid for non-overlapping inclusions as assumed here.
For the extension of these equations to 3D ellipsoidal inclusions, see Seibert~et~al. \cite{Seibert24}.

In practice, the truncated series
\begin{equation}
    \breve{S}_2(\ve{r}) = \sum_{(j,k) \in \mathcal{G}} \hat{S}_{2_{jk}} \,\mathrm{e}^{\mathrm{i} \boldsymbol{\omega}_{jk} \cdot \ve{r}}
\end{equation}
is evaluated on the same grid 
\begin{equation}
    \mathcal{G} = \left\{(j,k) \in \mathbb{Z}^2 \mid j \in [-N_1^\text{F}, N_1^\text{F}],\; k \in [-N_2^\text{F}, N_2^\text{F}]\right\},
\end{equation}
that is used for the computation of the Fourier coefficients, with $2N_1^\text{F} + 1$ and $2N_2^\text{F} + 1$ terms in the two spatial directions, respectively.
This allows us to use the inverse fast Fourier transform (IFFT).
Evaluating the series at the grid point $(u,v)$, located at position
\begin{equation}
    \ve{r}_{uv}
    = \frac{u}{2 N_1^\mathrm{F} + 1} L_1 \ve{e}_1 
    + \frac{v}{2 N_2^\mathrm{F} + 1} L_2 \ve{e}_2 ,
\end{equation}
yields the value
$
\breve{S}_{2_{uv}} 
= \breve{S}_2 \!\left(\ve{r}_{uv}\right).
$
The total number of grid points is given by $|\mathcal{G}| = (2 N_1^\text{F} + 1)(2 N_2^\text{F} + 1)$.
The SDF can be exactly computed from the TPC using the FFT, provided the same discretization is used, since no information is lost during the (inverse) FFT. 
Note that, while the equation $S_2(\ve{0}) = \phi$ holds, the corresponding value of the approximate two-point correlation $\breve{S}_2(\ve{0})$ deviates from the area fraction, with the deviation depending on the number of Fourier terms.
The TPC and its approximation are point-symmetric with respect to the origin, i.e., $S_2(\ve{r}) = S_2(-\ve{r})$, $\breve{S}_2(\ve{r}) = \breve{S}_2(-\ve{r})$, and $\breve{S}_{2_{uv}} = \breve{S}_{2_{-u,-v}}$.
The same symmetry holds for the SDF: $| \hat{\mathcal{I}}_{jk} |^2 = | \hat{\mathcal{I}}_{-j,-k} |^2$.
Therefore, two cells that are translated or rotated by \SI{180}{\degree} have the same TPC and SDF.
Consequently, the number of independent entries
$\breve{S}_{2_{uv}}$ and $| \hat{\mathcal{I}}_{jk} |^2$
is given by $\frac{\left(2 N_1^\text{F} + 1\right) \left(2 N_2^\text{F} + 1\right) + 1}{2}$.

\subsection{Optimization}
\label{sec:optimization}

For the microstructure generation, we distinguish between a reference cell, whose variables are denoted by $(\cdot)^\text{ref}$, and the generated cell, whose variables carry no superscript.
The reference cell is at least as large as the generated cell, i.e., $L^\text{ref}_1 \ge L_1$ and $L^\text{ref}_2 \ge L_2$.
All parameters of the reference cell are constant, whereas the particle radii $\underline{R}$ and center points $\underline{\ve{p}}$ of the generated cell are to be determined.

The objective is to generate a microstructure that minimizes the TPC loss
\begin{equation}
	\mathcal{L}^\text{TPC}\big(\underline{\ve{p}}\big) =
	\sum\limits_{(u,v) \in \mathcal{G}}
	\left( \breve{S}_{2_{uv}}\big(\underline{R},\underline{\ve{p}}\big) - \breve{S}_{2_{uv}}^\text{ref} \right)^2
\end{equation}
while maintaining a minimum particle distance.
To compute the TPC loss, the reference TPC must be evaluated on the same spatial grid $\mathcal{G}$ as the generated cell.
Equal grid indices then correspond to approximately equal spatial coordinates, provided that ${N^\text{F,ref}_{(\alpha)}}/{N^\text{F}_{(\alpha)}} \approx {L^\text{ref}_{(\alpha)}}/{L_{(\alpha)}}$, so that only the inner grid points are compared and the long-range information of the reference is neglected.

The minimum distance condition is expressed by the minimum-distance loss
\begin{equation}
	\mathcal{L}^\text{d}(\underline{R},\underline{\ve{p}}) =
	\sum_{l=1}^{N^\text{incl}} \sum_{q=l+1}^{N^\text{incl}}
	\op{ReLU}\left( d_{lq}^\text{pen} \right)^2 ,
\end{equation}
which is to be minimized, with $\op{ReLU}(x) = \op{max}(0,x)$ and the pairwise distance penalty
\begin{equation}
	d_{lq}^\text{pen} = d_\text{min} + R_l + R_q - \frob{\ve{d}(\ve{p}_l, \ve{p}_q)} \, ,
	\label{eq:loss_min_distance}
\end{equation}
where $d_\text{min}$ denotes the minimum particle distance.
The pairwise periodic distance vector between inclusions $l$ and $q$ is given by
\begin{equation}
	\ve{d}(\ve{p}_l, \ve{p}_q) = \sum_{\alpha=1}^{2} 
	\left(
	p_{l\alpha} - p_{q\alpha} - L_{(\alpha)} \left\lfloor \frac{p_{l\alpha} - p_{q\alpha}}{L_{(\alpha)}} + \frac{1}{2} \right\rfloor 	
	\right)
	\ve{e}_{\alpha} \, ,
\end{equation}
where the index $\alpha \in \{1,2\}$ denotes the direction and $\left\lfloor \circ \right\rfloor$ stands for the floor function.

Following Seibert~et~al.~\cite{Seibert24}, we use a two-step procedure for the microstructure generation, as shown in Algorithm~\ref{alg:microstructure_generation}:
First, we obtain a prescribed number of radii $\underline{R}$.
In the monodisperse case, all particles are assigned the area-equivalent mean radius of the reference microstructure, whereas in the polydisperse case, the radii are sampled from the reference microstructure.
In the following, the radii are considered constant.
The cell size is chosen such that the area fraction is identical and the ratios of the side lengths to the respective Fourier grid resolutions are similar.

Secondly, we determine the particle center points
\begin{equation}
	\underline{\ve{p}}^{\mathrm{opt}} = \underset{\underline{\ve{p}} \in \Omega^{\mathrm{cell}}}{\mathrm{argmin}} \, \mathcal{L}^\text{TPC}\big(\underline{\ve{p}}\big)
\end{equation}
that minimize the descriptor loss $\mathcal{L}^\text{TPC}$, whose derivative is obtained via the chain rule as
\begin{equation}
	\pd{\mathcal{L}^\text{TPC}\big(\underline{\ve{p}}\big)}{\ve{p}_l} =
	\sum\limits_{(u,v) \in \mathcal{G}}
	2 \left( \breve{S}_{2_{uv}}\!\big(\underline{\ve{p}}\big) - \breve{S}_{2_{uv}}^\text{ref} \right)
	\pd{\breve{S}_{2_{uv}}}{\ve{p}_l} .
\end{equation}
To calculate the derivative of the TPC
\begin{equation}
	\pd{\breve{S}_{2}(\ve{r})}{\ve{p}_l} = 
	\sum\limits_{(j,k) \in \mathcal{G}}
	\pd{|\hat{\mathcal{I}}_{jk}|^2}{\ve{p}_l}
	\,\mathrm{e}^{\mathrm{i} \boldsymbol{\omega}_{jk} \cdot \ve{r}} ,
\end{equation}
we apply an IFFT to the derivative of the SDF, which is given by
\begin{equation}
\label{eq:d_Ijk_d_pl}
\begin{split}
\pd{|\hat{\mathcal{I}}_{jk}|^2}{\ve{p}_l} =
2 \boldsymbol{\omega}_{jk}
\Bigl(
    &\op{Re}\left(\hat{\mathcal{I}}_{jk}\right)
    \op{Im}\left(\hat{\mathcal{I}}^\text{incl}_{jk}\left(R_l, \ve{p}_l\right)\right) \\
    - &\op{Im}\left(\hat{\mathcal{I}}_{jk}\right)
    \op{Re}\left(\hat{\mathcal{I}}^\text{incl}_{jk}\left(R_l, \ve{p}_l\right)\right)
\Bigr).
\end{split}
\end{equation}
The descriptor loss and its derivative are then used by an optimizer to compute the update of the particle center points $\Delta \underline{\ve{p}}$.

Each descriptor optimization step is preceded by the enforcement of the minimum-distance constraint.
After computing the corresponding derivative as
\begin{equation}
	\pd{\mathcal{L}^\text{d}}{\ve{p}_l} =
	-2 \sum\limits_{\substack{q=1 \\ q \neq l}}^{N^\text{incl}}
	\op{ReLU}\left(d_{lq}^\text{pen}\right) \, \frac{\ve{d}(\ve{p}_l, \ve{p}_q)}{\frob{\ve{d}(\ve{p}_l, \ve{p}_q)}} ,
\end{equation}
we obtain the update $\Delta\ve{p}_l = - \alpha \pd{\mathcal{L}^\text{d}}{\ve{p}_l}$, which minimizes $\mathcal{L}^\text{d}$ by gradient descent with the step size $\alpha = 0.501$.
After the optimization, the particles are mapped to the cell domain $\Omega^\text{cell}$ taking the periodicity into account.

\begin{algorithm}
	\caption{Proposed microstructure generation algorithm}
	\label{alg:microstructure_generation}
	\begin{algorithmic}[1]
		
		\Require reference cell (parameterized by $\underline{R}^{\text{ref}}, \underline{\ve{p}}^{\text{ref}}$, $L^\text{ref}_1$, $L^\text{ref}_2$), number of particles $N^\text{incl}$ in the cell to be generated, resolution ratio 
        $\varrho_{(\alpha)}$
		\State $\phi^\text{ref} \gets \frac{\sum_l \pi (R_l^\text{ref})^2}{L_1^\text{ref} L_2^\text{ref}}$

		\If{monodisperse}
			\State $R_q \gets \sqrt{\frac{\sum_l (R_l^\text{ref})^2}{N^{\text{incl}}_{\text{ref}}}}$
		\ElsIf{polydisperse}
			\State $\underline{R} \gets $ draw $N^\text{incl}$ random values from $\underline{R}^\text{ref}$
		\EndIf

		\State $L_{1,2} \gets \sqrt{\frac{\sum_l \pi R_l^2}{\phi^\text{ref}} \frac{L_{1,2}^\text{ref}}{L_{2,1}^\text{ref}}}$
		\Comment{ensure $\phi = \phi^\text{ref}$ and $\frac{L_1}{L_2} = \frac{L_1^\text{ref}}{L_2^\text{ref}}$}

		\State $N^\text{F}_{(\alpha)} \gets \op{round} \left(\varrho_{(\alpha)} \sqrt{N^\text{incl}}\right)$ 
		\State $N^\text{F,ref}_{(\alpha)} \gets \op{round} \left(N^\text{F}_{(\alpha)} \frac{L^\text{ref}_{(\alpha)}}{L_{(\alpha)}} \right)$
		\Comment{ensure $\frac{N^\text{F,ref}_{(\alpha)}}{N^\text{F}_{(\alpha)}} \approx \frac{L^\text{ref}_{(\alpha)}}{L_{(\alpha)}}$}		

        \State initialize $\underline{\ve{p}}$
		\Comment{by random placement}

		\For{$i \gets 1$ \textbf{to} $N^\text{iter}$}
		
		\State $\underline{\ve{p}} \gets \underset{\underline{\ve{p}} \in (\Omega^{\mathrm{cell}})^{N^{\mathrm{incl}}}}{\mathrm{argmin}} \, \mathcal{L}^\text{d}\big(\underline{\ve{p}}\big)$
		\Comment{enforce min. distance}

		\State compute $\mathcal{L}^\text{TPC}$, $\pd{\mathcal{L}^\text{TPC}}{\ve{p}_l}$

		\If{$i = N^\text{iter}$ or $\mathcal{L}^\text{TPC} < \varepsilon_\text{tol}$}
			\State \textbf{break}
		\EndIf

		\State compute $\Delta \underline{\ve{p}}$ with optimizer

		\State $\underline{\ve{p}} \gets \underline{\ve{p}} + \Delta \underline{\ve{p}}$
		\Comment{apply descriptor step}

		\EndFor
	
	\State \Return generated cell (parameterized by $\underline{R}, \underline{\ve{p}}$, $L_1$, $L_2$)
\end{algorithmic}
\end{algorithm}

\section{Finite strain magneto-hyperelasticity}
\label{sec:Finite_strain_magneto-hyperelasticity}

Once unit cells are generated, their effective response under coupled magneto-mechanical loading is compared to that of the reference cell. 
To this end, the Maxwell equations and mechanical balance laws are solved for a prescribed macroscopic load state and corresponding boundary conditions, taking into account the constitutive behavior of the individual phases. 
The resulting microscale fields are then averaged to obtain effective macroscale quantities, enabling a comparison between generated and reference area elements.

In the following, we provide the general three-dimen\-sional equations.
For the computations in Sec.~\ref{sec:homogenization_results}, we assume a plane strain state and two-dimensional magnetic quantities in the $X_1$-$X_2$ plane. 
For further details, we refer to the related work \cite{Kalina24}.

\subsection{Kinematics}

Let $\mathcal{B}_0 \subset \mathbb{R}^3$ denote the reference configuration of a deformable body at time $t_0$, and $\mathcal{B} \subset \mathbb{R}^3$ its current configuration at time $t \in \mathcal{T} := \{ t \in \mathbb{R} \mid t \ge t_0 \}$. 
The deformation map $\ve{\varphi} : \mathcal{B}_0 \times \mathcal{T} \to \mathcal{B},\ (\ve{X},t) \mapsto \ve{x}(\ve{X},t)$ assigns to each referential point $\ve{X} \in \mathcal{B}_0$ its current position $\ve{x} \in \mathcal{B}$.
In the following, the dependency of field variables on $\ve{X}$ and $t$ is omitted.
The displacement vector $\ve{u} = \ve{x} - \ve{X}$ and velocity $\ve{v} = \dot{\ve{u}}$ with the material time derivative $\dot{(\circ)}$ are given as shown.
The deformation gradient $\te{F} = (\Grad \ve{\varphi})^\text{T} = \te{1} + (\Grad \ve{u})^\text{T}$ may be split into the multiplicative form $\te{F} = \te{F}^\text{vol} \cdot \te{F}^\text{iso}$, where the volumetric deformation gradient is given by $\te{F}^\text{vol} = J^{\frac{1}{3}} \te{1}$ and the isochoric deformation gradient by $\te{F}^\text{iso} = J^{-\frac{1}{3}} \te{F}$, with the Jacobian determinant $J = \op{det}(\te{F}) > 0$.
A common deformation measure that is independent of rigid body motions is the right Cauchy--Green deformation tensor $\te{C} = \te{F}^\text{T} \cdot \te{F}$.
The square roots of its eigenvalues are the principal stretches $\lambda_l \in \mathbb{R}_{>0}, l \in \{1,2,3\}$.
The linearization of the Green--Lagrange strain tensor $\te{E} = \frac{1}{2}\left(\te{C} - \te{1}\right)$ yields the linearized strain tensor $\te{\upvarepsilon} = \frac{1}{2}\left(\grad \ve{u} + (\grad \ve{u})^\text{T}\right)$.
For a more comprehensive treatment of kinematics in continuum theory, the reader is referred to the textbooks \cite{Ciarlet88,Silhavy97,Holzapfel2002}.

\subsection{Maxwell and mechanical balance equations}

The magnetic behavior of a quasistatic system without electric fields and currents is governed by the two Maxwell equations 
\begin{equation}
    \label{eq:maxwell}
    \div \ve{b} = 0 
	\quad \text{and} \quad 
    \rot \ve{h} = \ve{0} .
\end{equation}
For the jump conditions over material surfaces, we refer to a previous publication \cite{Roth2026}.
The magnetic induction $\ve{b} = \mu_0 (\ve{h} + \ve{m})$ and magnetic field strength $\ve{h}$ are linked by the magnetization $\ve{m}$ and the vacuum permeability $\mu_0 = 4\pi\cdot10^{-7} \,\si{\newton\per\ampere\squared}$.
To account for magneto-mechanical coupling, the ponderomotive force density $\ve{f}^\text{pon} = \ve{m} \cdot \grad \ve{b}$, ponderomotive torque density $\ve{c}^\text{pon} = \ve{m} \times \ve{b}$, and ponderomotive power density $p^\text{pon} = \ve{f}^\text{pon} \cdot \ve{v} - \ve{m} \cdot \dot{\ve{b}}$, formulated here according to the field-matter interaction model of de Groot and Suttorp \cite{degroot69}, need to be considered. 
The effects of the first two of these quantities are expressed using the ponderomotive stress tensor
\begin{equation}
    \te{\upsigma}^\text{pon} = \frac{1}{\mu_0}\left( \ve{b} \otimes \ve{b} - \frac{1}{2}\frob{\ve{b}}^2 \te{1} \right) + (\ve{b} \cdot \ve{m}) \te{1} - \ve{m} \otimes \ve{b},
\end{equation}
which satisfies the conditions
$\div (\te{\upsigma}^\text{pon})^\text{T} = \ve{f}^\text{pon}$ and $\op{skw}(\te{\upsigma}^\text{pon}) = - \frac{1}{2} \mathfrak{e} \cdot \ve{c}^\text{pon}$ with the permutation pseudotensor $\mathfrak{e} = \sum_{\alpha\beta\gamma} e_{\alpha\beta\gamma} \ve{e}_\alpha \otimes \ve{e}_\beta \otimes \ve{e}_\gamma$, where $e_{\alpha\beta\gamma}$ is the Levi--Civita symbol \cite{degroot69,Kalina20-1}.
By defining the total stress tensor $\te{\upsigma}^\text{tot} = \te{\upsigma} + \te{\upsigma}^\text{pon}$ as the sum of the mechanical stress tensor $\te{\upsigma}$ and the ponderomotive stress tensor, the balance of linear and angular momentum without mechanical volume forces
$\div \te{\upsigma}^\text{tot} = \ve{0}$ and $\op{skw}(\te{\upsigma}^\text{tot}) = \te{0}$
may be formulated in a compact form.
The referential field variables
$\ve{B} = \ve{b} \cdot \op{cof}(\te{F})$,
$\ve{H} = \ve{h} \cdot \te{F}$, and
$\Ptot = \sigtot \cdot \op{cof}(\te{F})$
are obtained through pull-back operations \cite{Dorfmann04,Kankanala04}, with $\Ptot$ denoting the total first Piola--Kirchhoff stress tensor.

\subsection{Constitutive equations}

For a magnetohyperelastic solid with the total Helmholtz free energy density 
\begin{equation}
    \Psi : \mathcal{GL}^+(3) \times \mathbb{R}^3 \to \mathbb{R}, \ \left(\te{F}, \ve{B}\right) \mapsto \Psi(\te{F}, \ve{B}),
\end{equation}
where $\mathcal{GL}^+(3)$ denotes the set of invertible second-order tensors with a positive determinant,
the definitions
\begin{equation}
    \label{eq:Ptot_Psi_F_H_Psi_B}
        \Ptot = \pd{\Psi}{\te{F}} 
        \quad \text{and} \quad 
        \ve{H} = \pd{\Psi}{\ve{B}}
\end{equation}
ensure that the Clausius--Duhem inequality is satisfied \cite{Coleman1963}.
Since the matrix is non-magnetizable and the particles are effectively rigid, the magneto-mechanical coupling within each constituent of the MRE is negligible. 
Accordingly, the total Helmholtz free-energy density decomposes additively
\begin{equation}
    \label{eq:psi_mech_psi_mag_psi_vac}
    \Psi(\te{F},\ve{B}) 
    = \Psi^\text{iso}(\te{F}^\text{iso}) 
    + \Psi^\text{vol}(J) 
    + \Psi^\text{mag}(\te{F},\ve{B})
    + \Psi^\text{vac}(\te{F},\ve{B})
\end{equation}
without requiring a coupling term. 
A volumetric contribution of
\begin{equation}
    \label{eq:psi_vol}
    \Psi^\text{vol} = 
    \frac{\kappa}{4}\left(J^2 - 2 \op{ln}(J) - 1\right) ,
\end{equation}
with the bulk modulus $\kappa = \frac{2}{3} \mu \frac{1+\nu}{1 - 2 \nu}$
is chosen for both matrix and particles with the respective shear modulus $\mu$ and Poisson's ratio $\nu$.
The material parameters $\mu_l$ and $\alpha_l$ of the matrix are shown in Tab.~\ref{tab:material_params_matrix}.

\begin{table*}[]
    \begin{center}
        \small
        \caption{Material parameters of the matrix phase, including the initial shear modulus $\mu^\text{M}$ and Poisson's ratio $\nu^\text{M}$, taken from Kalina \cite{Kalina21Diss} with a silicone-oil ratio of 20\%. The Poisson's ratio has been increased.}
        \begin{tabular}{ c c c c c c c c }
        \hline
            $\mu_1 / \si{\kilo\pascal}$ & $\mu_2 / \si{\kilo\pascal}$ & $\mu_3 / \si{\kilo\pascal}$ & $\alpha_1$ & $\alpha_2$ & $\alpha_3$ & $\mu^\text{M}$ / \si{\kilo\pascal} & $\nu^\text{M}$ \\ 
            \hline
            -11.80 & 12.45 & $4.59\cdot10^{-5}$ & -6.68 & 2.09 & 18.34 & 52.40 & 0.495 \\
            \hline
        \end{tabular}
    \label{tab:material_params_matrix}
    \end{center}
\end{table*}

For the matrix, we select the Ogden model \cite{Ogden97} and prescribe a vanishing magnetization energy, i.e.,
\begin{equation}
    \label{eq:psi_Matrix}
    \Psi^\text{iso,M} = 
    \sum_{l=1}^3 
    \left(
    \frac{\mu_l}{\alpha_l}
    \left( (\lambda_1^\text{iso})^{\alpha_l} + 
    (\lambda_2^\text{iso})^{\alpha_l} + 
    (\lambda_3^\text{iso})^{\alpha_l} - 3 \right)
    \right)
\end{equation}
and
$
\Psi^\text{mag,M} = 0 .
$

For the particles, the Neo-Hookean model \cite{Ciarlet88} and the Langevin model \cite{Danas17}, given by
\begin{equation}
    \label{eq:psi_Particle}
    \Psi^\text{iso,P} = 
    \frac{\mu^\text{P}}{2} \left(\op{tr}{\te{C}^\text{iso}} - 3 \right)
\end{equation}
and
\begin{equation}
    \Psi^\text{mag,P} =
    \begin{cases}
        \displaystyle
        \frac{\mu_0 m_\text{s}^2}{3\chi}
        \ln\left(
        \frac{\frac{3\chi}{\mu_0m_\text{s}}J^{-\frac12}
        \frob{\te{F}\cdot\ve{B}}}
        {\sinh\left(
        \frac{3\chi}{\mu_0m_\text{s}}J^{-\frac12}
        \frob{\te{F}\cdot\ve{B}}
        \right)}
        \right),
        & \ve{B}\neq\ve{0}, \\[3ex]
        0,
        & \ve{B}=\ve{0},
    \end{cases}
\end{equation}
respectively, are chosen with the magnetic parameters \\ $\chi = 0.9$ and $m_\text{s} = \SI{1000}{\kilo\ampere\per\meter}$ in accordance with Kalina~et~al. \cite{Kalina24}, together with the mechanical parameters $\mu^\text{P} = \SI{251.1}{\mega\pascal}$ and $\nu^\text{P} = 0.3$.
The vacuum free energy $\Psi^\text{vac} = \frac{1}{2\mu_0 J} \frob{\te{F} \cdot \ve{B}}^2$ is material-independent.

\subsection{Scale transition and variational formulation}

MRE samples exhibit a pronounced separation of scales between the microscopic structure and the macroscopic sample dimensions, which enables an efficient two-scale simulation of their behavior.
On the microscale, individual constituents, which have the minimum characteristic length $\mathscr{l}$, are fully resolved.
On the macroscale, where samples with characteristic lengths of $\bar{\mathscr{l}}$ are considered, the material is considered to be homogeneous.
The scale separation is given by the relation $\bar{\mathscr{l}} \gg \mathscr{l}$ \cite{Schroeder13}.
To determine the effective material behavior, it is sufficient to consider an area element which is large enough to be representative, called RVE.
Then, the macroscopic quantities
\begin{equation}
	\label{eq:homogenization_F_P_B_H}
	\bar{\te{F}} = \ho{\te{F}}_0 \  , 
	\  \bar{\te{P}}^\text{tot} = \ho{\te{P}^\text{tot}}_0 \  , 
	\  \ve{\bar{B}} = \ho{\ve{B}}_0 \  \text{and}
	\  \ve{\bar{H}} = \ho{\ve{H}}_0
\end{equation}
are calculated using the averaging operator
\begin{equation}
    \ho{\circ}_0 = \frac{1}{|\mathcal{B}_0^\text{RVE}|} \int_{\mathcal{B}_0^\text{RVE}} \circ \ \dd V .
\end{equation}
All other macroscopic quantities, marked by $\bar{(\circ)}$, are calculated not by averaging the respective microscopic quantity, but by applying the equations in this section on the quantities from Eq.~\eqref{eq:homogenization_F_P_B_H} \cite{Chatzigeorgiou14}.
The deformation gradient and magnetic induction are additively split into a macroscopic part and a fluctuation part, i.e.,
\begin{equation}
    \label{eq:F_B_split}
    \te{F} = \bar{\te{F}} + (\Grad \tilde{\ve{u}})^\text{T} 
    \quad \text{ and } \quad
    \ve{B} = \ve{\bar{B}} + \Rot \tilde{\ve{A}} ,
\end{equation}
where $\tilde{\ve{A}} : \mathcal{B}_0 \times \mathcal{T} \to \mathbb{R}^3, (\ve{X},t) \mapsto \tilde{\ve{A}}(\ve{X}, t)$, denotes the magnetic vector potential fluctuation.
Fluctuations are marked by $\tilde{(\circ)}$. 
One possible set of boundary conditions to satisfy the Hill--Mandel condition \cite{Hill63,Chatzigeorgiou14}
$\ho{\te{P}^\text{tot} : \dot{\te{F}}{}^\text{T} + \ve{H} \cdot \dot{\ve{B}}}_0 = \bar{\te{P}}^\text{tot} : \dot{\bar{\te{F}}}^\text{T} + \ve{\bar{H}} \cdot \dot{\bar{\ve{B}}}$
are the periodic boundary conditions \cite{Zabihyan18,Zabihyan20}
\begin{equation}
        \label{eq:periodic_bc}
            \jump{\tilde{\ve{u}}} = \ve{0} , \ 
            \jump{\tilde{\ve{A}}} \times \ve{N} = \ve{0} , \
            \jump{\te{P}^\text{tot}} \cdot \ve{N} = \ve{0}, \ \text{and} \
            \jump{\ve{H}} \times \ve{N} = \ve{0} ,
\end{equation}
with the operator $\jump{(\circ)} = (\circ)^+ - (\circ)^-$ specifying the difference between variables on opposite sides of the RVE.

To determine the behavior when the macroscopic loadings $\bar{\te{F}}$ and $\bar{\ve{B}}$ are prescribed, the primary variables $\tilde{\ve{u}}, \tilde{\ve{A}}$,
dilatation $\theta$, and pressure $p$ are used.
By utilizing dilatation and pressure as primary variables, volumetric locking for quasi-incompressible materials is avoided \cite{Simo85,Okada09,Polukhov18}.
To obtain a suitable solution to the boundary value problem, it is formulated as the extremum value problem 
\begin{equation}
	(\tilde{\ve{u}}, \tilde{\ve{A}}, \theta, p) =
	\op{arg} \left(
		\underset{\tilde{\ve{u}}}{\op{inf}} \ 
		\underset{\tilde{\ve{A}}}{\op{inf}} \ 
		\underset{\theta}{\op{inf}} \ 
		\underset{p}{\op{sup}} \ 
		\Pi(\tilde{\ve{u}}, \tilde{\ve{A}}, \theta, p)
	\right) ,
\end{equation}
which needs to be solved numerically with the potential
\begin{equation}
    \label{eq:Pi_micro}
    \begin{split}
        \Pi(\tilde{\ve{u}}, \tilde{\ve{A}}, \theta, p)
        &= \int_{\mathcal{B}_0}
        \Bigl(
            \Psi^\text{iso}(\te{F}^\text{iso})
            + \Psi^\text{vol}(\theta) \\
        &
            + \Psi^\text{mag}(\te{F}, \ve{B})
            + \Psi^\text{vac}(\te{F}, \ve{B}) \\
        &
            + p (J - \theta)
            + \frac{\kappa^\text{reg}}{2}\frob{\tilde{\ve{A}}}^2
        \Bigr)
        \, \dd V .
    \end{split}
\end{equation}
For more details on the variational formulation we refer to the related works \cite{Polukhov18,Roth2026}.
The regularization term with the small parameter $\kappa^\text{reg}$ ensures the well-posedness of the problem \cite{Reitzinger02,ZaglmayrDiss,Gebhart24}.

\section{Examples}
\label{sec:Numerical_examples}

To validate the microstructure generation procedure proposed in Sec.~\ref{sec:Microstructure_generation}, we generate cells using a reference cell obtained from a photograph. We compare the generated cells with the reference cell in terms of their simulated effective initial stiffness tensor, their initial magnetic susceptibility tensor, and their response under a coupled magneto-mechanical load case.

\subsection{Determination of the reference microstructure}
\label{sec:Determination_of_the_reference_microstructure}

The reference microstructure is obtained in two steps.
First, a microstructure photograph is acquired experimentally.
Subsequently, individual particles are segmented and parameterized from the image.

\subsubsection{Experimental setup}
\label{sec:Experimental_setup}
In order to obtain a reference microstructure, particles in a matrix need to be photographed. The sample is imaged in a transmission using a Photron fast camera Mini AX 200 equipped with a Navitar Resolv4K Zoom lens and an iris diaphragm positioned in front of it. 
The sample is illuminated by a KL 2500 LED SCHOTT Light Source.
With the iris diaphragm we increase the depth of field. 

\begin{figure}
    \centering
    \includegraphics[width=0.9\linewidth]{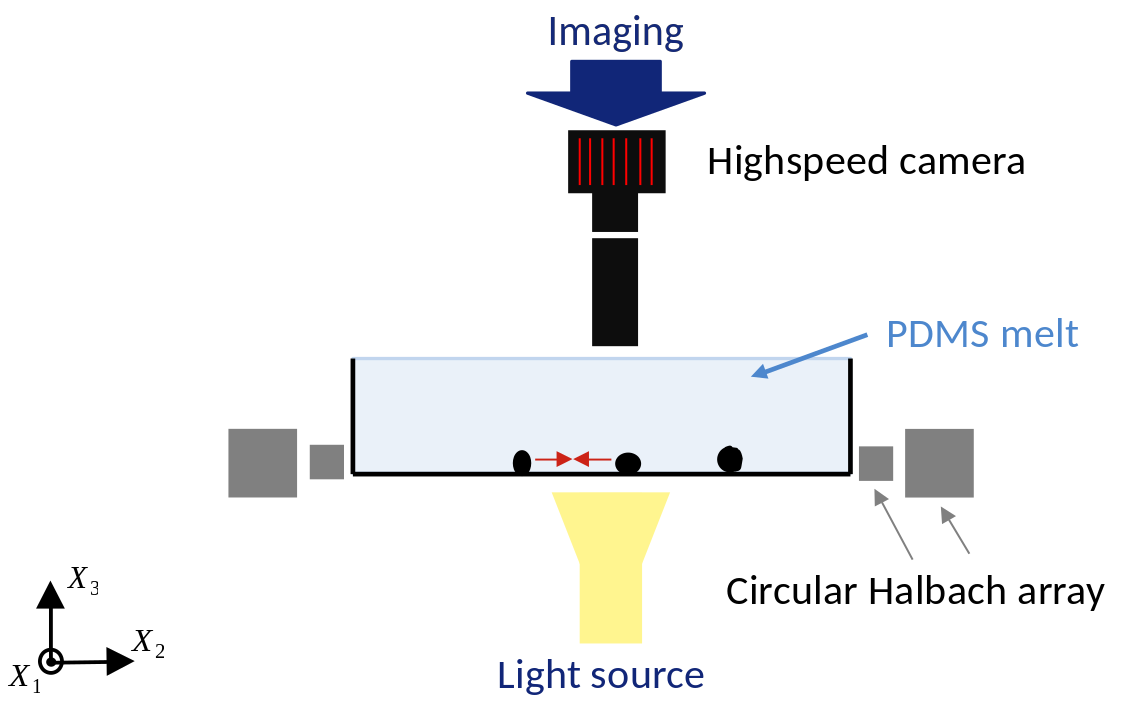}
    \caption{Scheme of the experimental setup used to observe the particle chain formation from the top.}
	\label{fig:experimental_setup}
\end{figure}

The used magnetically soft nickel particles are approximately spherical and are supplied by Thermo Scientific. These particles are sieved to a $-100/+325$ mesh, corresponding to a particle size of approximately \SIrange{44}{150}{\micro\meter}. To create an MRE, the particles are embedded in a matrix consisting of PDMS melt from the Kurt Obermeier GmbH with a viscosity of \SI{300000}{cSt}. For this purpose, the PDMS melt is first poured into a cylindrical glass Petri dish with a diameter of \SI{3}{\centi\metre}.
After the particles are finely dispersed over this PDMS melt, the sample is degassed for \SI{30}{\minute} in a vacuum oven at \SI{50}{\degreeCelsius} in order to remove air bubbles and to ensure that all particles settle in a single layer at the bottom of the sample. Afterwards, the sample is cooled down to room temperature over a period of \SI{30}{\minute}. The sample is then placed in a magnetic field of \SI{216}{\milli\tesla}, applied in the $X_2$-direction using double-stacked circular Halbach arrays \cite{Huang2016}, causing the particles to form chains. A schematic of the setup is shown in Fig.~\ref{fig:experimental_setup}. The chain formation itself occurs over \SI{1}{\minute}. Thereafter, no additional movement of the particles is observed. This final state is then shown in Fig.~\ref{fig:particle_segmentation_Original}. Note that no crosslinking has yet occurred.

\subsubsection{Particle segmentation}
\label{sec:Particle_segmentation}
\begin{figure*}[]
    \centering
    \begin{subfigure}{0.32\textwidth}
        \centering
        \fbox{\includegraphics[width=0.99\linewidth]{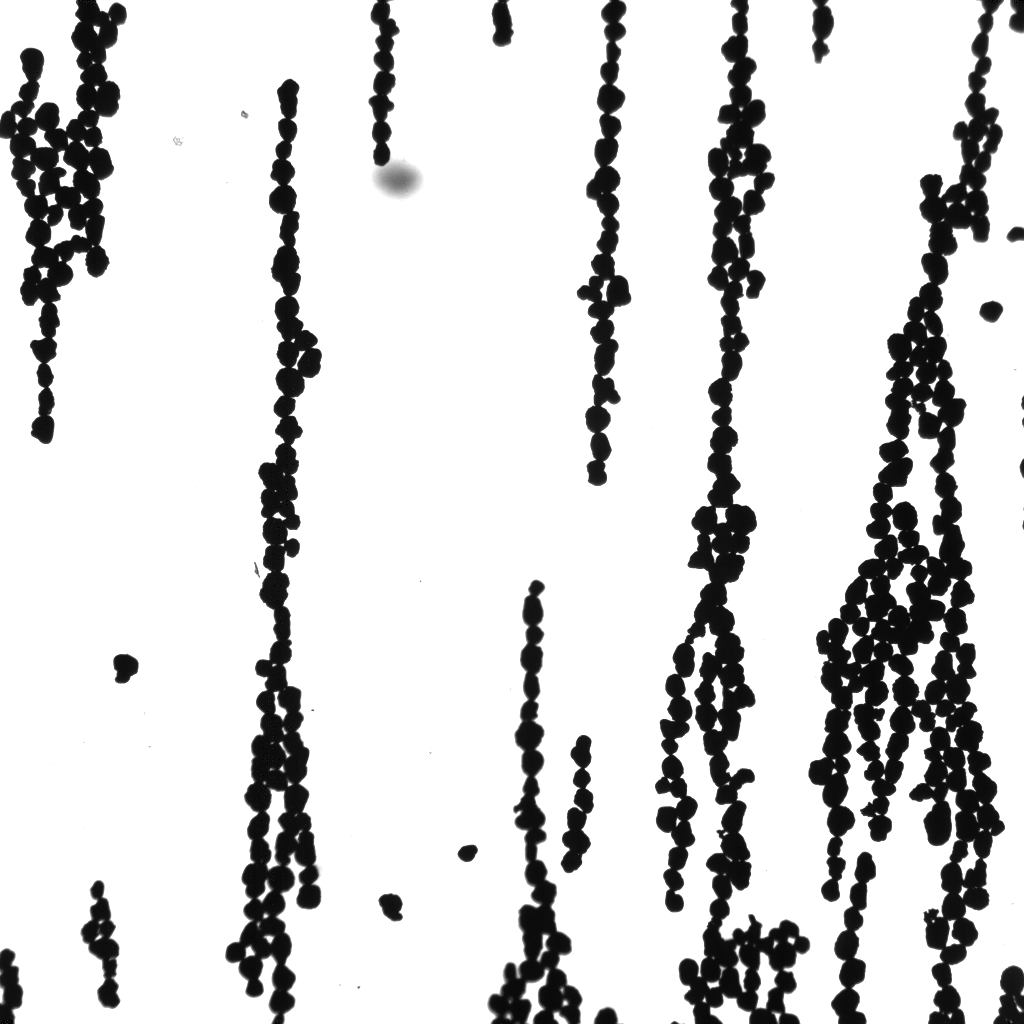}}
        \caption{Original photograph of the particles in the PDMS melt.}
		\label{fig:particle_segmentation_Original}
    \end{subfigure}
    \hfill
    \begin{subfigure}{0.32\textwidth}
        \centering
        \fbox{\includegraphics[width=0.99\linewidth]{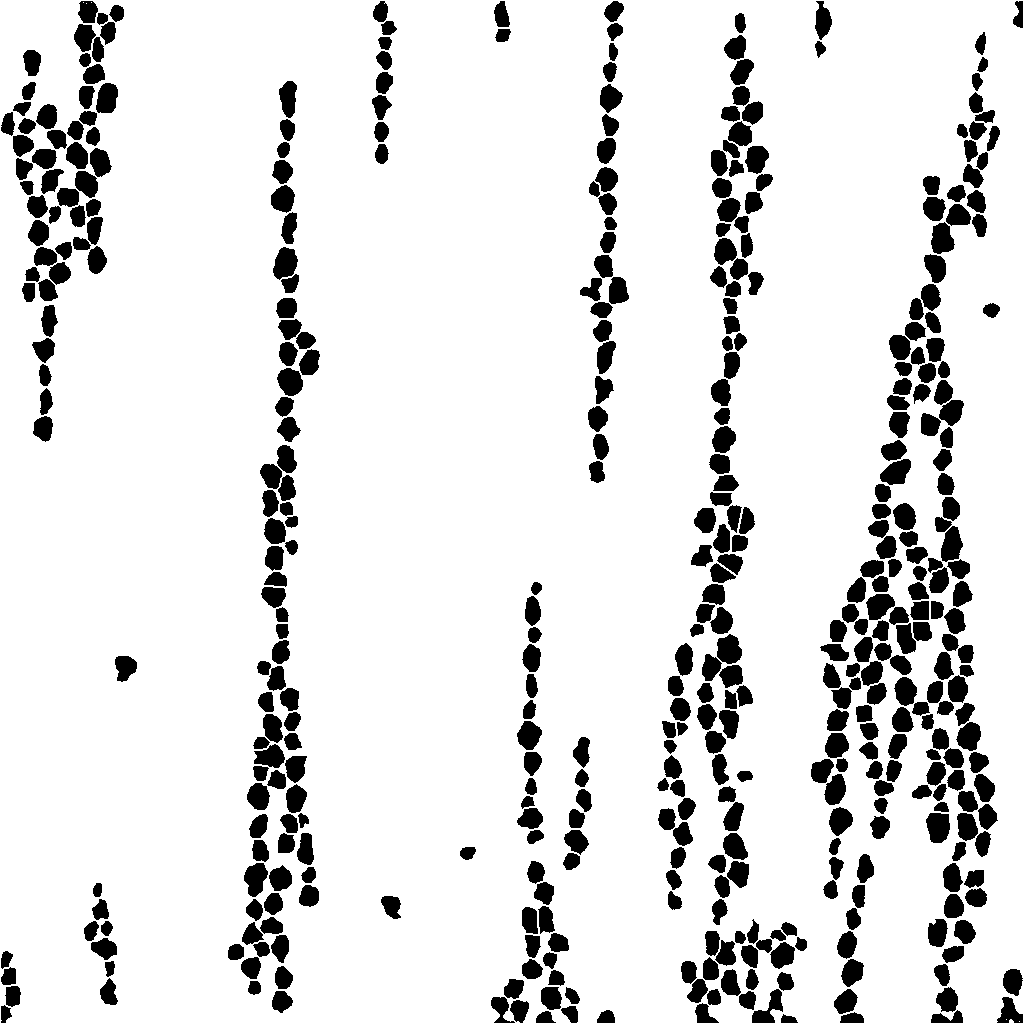}}
        \caption{Microstructure with segmented particles.\\~}
		\label{fig:particle_segmentation_Segmentation}
    \end{subfigure}
    \hfill
    \begin{subfigure}{0.32\textwidth}
        \centering
        \fbox{\includegraphics[width=0.99\linewidth]{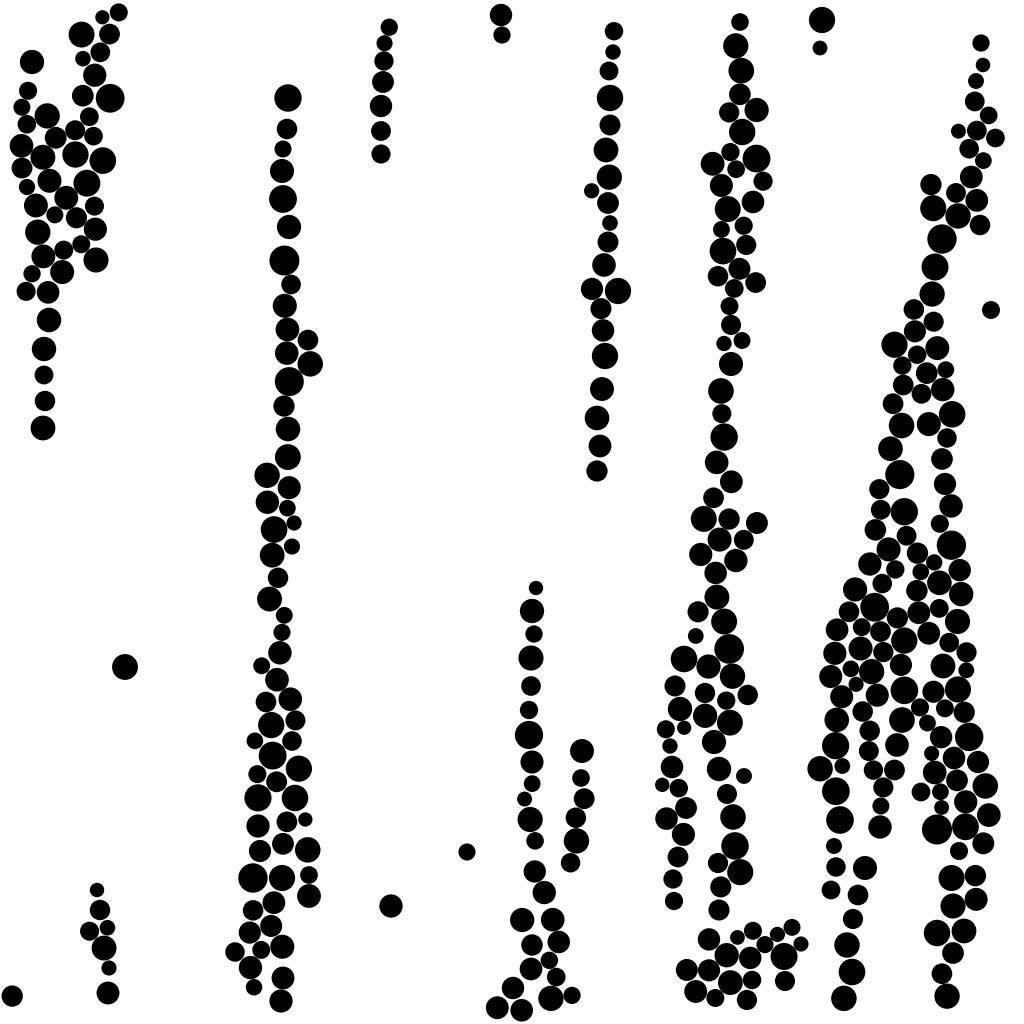}}
        \caption{Parameterized microstructure with particles approximated by 429 circles.}
		\label{fig:particle_segmentation_circles}
    \end{subfigure}
    \caption{Particle segmentation procedure for obtaining the parameterized reference microstructure from a photograph of nickel particles in a liquid PDMS matrix.} 
	\label{fig:particle_segmentation}
\end{figure*}

To obtain a parameterized reference cell, inclusions must be extracted from a raster image.
For this purpose, we employ the Python libraries OpenCV and scikit-image.
The reference image with $1024 \times 1024$ pixels shown in Fig.~\ref{fig:particle_segmentation_Original} is obtained as described in Sec.~\ref{sec:Experimental_setup}.
In the following, we use dimensionless spatial units in which the pixels are of size 1.

We begin the image processing by applying a Contrast Limited Adaptive Histogram Equalization (CLAHE) algorithm to increase the contrast. 
The resulting 8-bit grayscale image is converted to a black-and-white image by thresholding at an intensity value of 60.
To identify the particles, a watershed algorithm \cite{Vincent1991} with anisotropic seed distance filtering is used.
Very small regions with an area of less than 100 pixels are merged, while large connected regions with a low ratio of area to convex-hull area, indicating pronounced concavity, are split into separate particles as follows:
First, the point at which the minimum distance between the surface and its convex hull is largest is identified.
A line perpendicular to the surface is then constructed through this point.
The region is finally split into two particles along this line.
Very small regions with less than 100 pixels are omitted.
The segmented particles are shown in Fig.~\ref{fig:particle_segmentation_Segmentation}.

To parameterize the microstructure, we fit circles to the particles such that their centers of mass and areas are preserved. 
The circles may be very close to each other or even overlap, so a minimum distance must be enforced, which is chosen as $d_\text{min} = 0.2 \op{max}(\underline{R}) \approx 2.7$.
This minimum distance is important for stable FE simulations and is achieved by moving the particles such that the violation of the minimum distance, quantified by Eq.~\eqref{eq:loss_min_distance}, is minimized.
Particles lying on the boundary are removed, resulting in the cell with 429 particles shown in Fig.~\ref{fig:particle_segmentation_circles}.
To enforce periodicity, the cell is mirrored along the right and bottom boundaries, including the lower-right corner, resulting in a domain with side lengths twice the original size \cite{Grimm-Strele2021}. 
The final reference area element has 1716 particles and a particle area fraction of $\phi = 0.1188$.
A different microstructure with a higher particle area fraction is considered in App.~\ref{app:Microstructure_with_higher_area_fraction}.

\setlength{\fboxrule}{0.5pt}
\setlength{\fboxsep}{0pt}

\subsection{Microstructure generation}
\label{sec:Numerical_examples_Microstructure_generation}

\begin{figure*}[h]
    \centering
    \begin{subfigure}{0.32\textwidth}
        \centering
        \includegraphics[width=\linewidth]{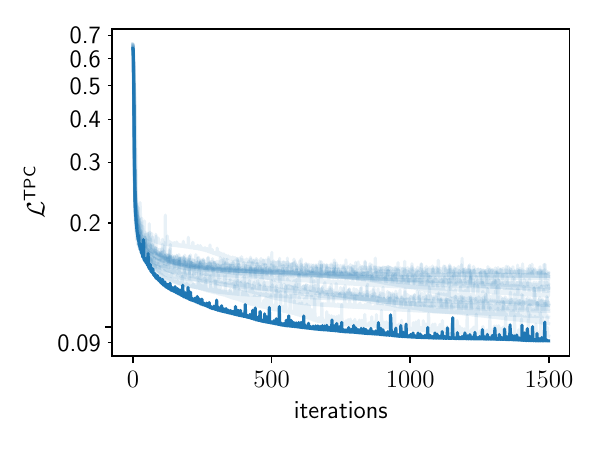}
        \caption{$N^\text{incl} = 400$}
		\label{fig:optimization_400}
    \end{subfigure}
    \begin{subfigure}{0.32\textwidth}
        \centering
        \includegraphics[width=\linewidth]{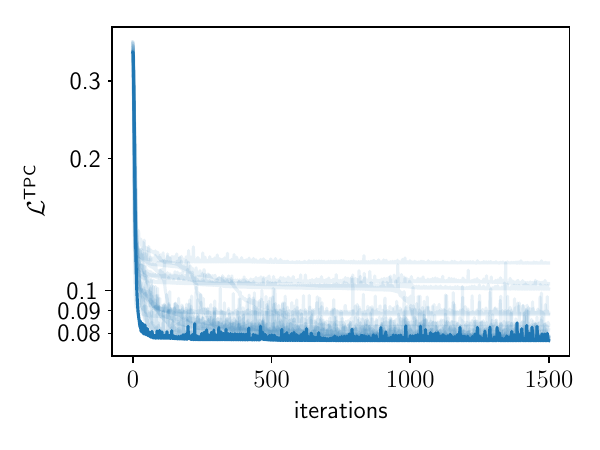}
        \caption{$N^\text{incl} = 64$}
		\label{fig:optimization_64}
    \end{subfigure}
    \begin{subfigure}{0.32\textwidth}
        \centering
        \includegraphics[width=\linewidth]{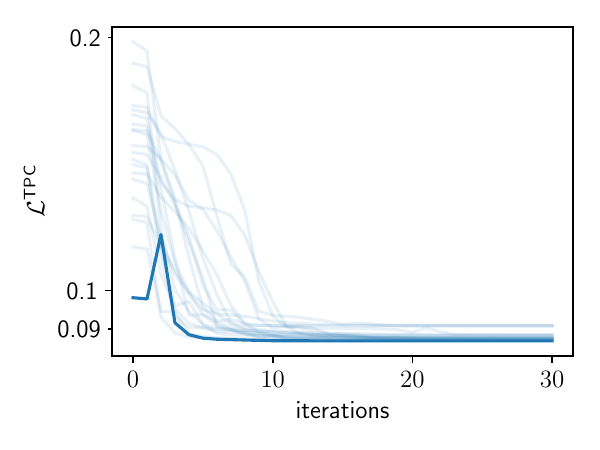}
        \caption{$N^\text{incl} = 4$}
		\label{fig:optimization_4}
    \end{subfigure}
    \caption{Loss during the generation of unit cells containing 400, 64, and 4 particles, respectively.
    For each size, the loss is shown for 20 different initializations.
    The optimization run with the lowest final loss is highlighted.
    For $N^\text{incl} = 4$, no further improvement occurs after 30 iterations.}
	\label{fig:optimization}
\end{figure*}

Utilizing the cell obtained from the image in Sec.~\ref{sec:Particle_segmentation} as the reference, we generate cells with a varying number of monodisperse particles $N^\text{incl} \in \{4, 9, 16, 36, 64, 256, 400\}$ as outlined in Sec.~\ref{sec:Microstructure_generation} by minimizing the descriptor loss.
Since the reference cell is of square shape, we set $L_1^\text{ref} = L_2^\text{ref}$.
The generated cells are also chosen to be of square shape, i.e., $L_1 = L_2$.
The resolution ratio 
$\varrho = 4$ 
is chosen for both the reference and the generated cells.
We have shown that increasing the resolution does not systematically reduce the differences between the effective properties examined in Sec.~\ref{sec:homogenization_results}.
As shown in App.~\ref{app:Generation_of_polydisperse_microstructures}, generating cells with polydisperse particles does not lead to a significantly lower error in the effective properties. 
We therefore generate cells with monodisperse particles.

Each of the 20 differently initialized cells is optimized over 1500 iterations using the optimization algorithm proposed by Malitsky and Mishchenko \cite{Malitsky2020}, implemented in Cython.
As shown in Fig.~\ref{fig:optimization}, for smaller numbers of particles, the loss no longer decreases after the first few iterations, whereas for area elements with more particles it still decreases gradually even after many iterations.
The generation of an area element containing 400 particles using the TPC takes approximately \SI{3}{\hour} on a single CPU core.
For smaller cells, the generation is considerably faster, with 64 particles requiring only about \SI{5}{\minute}.
In the following, we only consider the five generated cells with the lowest final loss for each respective particle count.

\begin{figure*}[]
    \centering
        \includegraphics[width=0.3\linewidth]{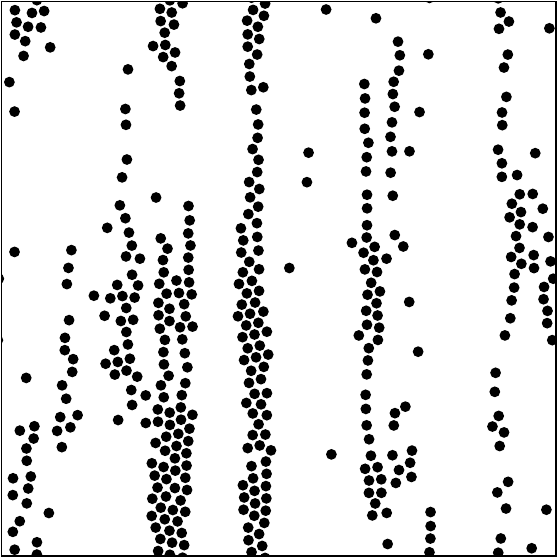}
        \includegraphics[width=0.24\linewidth]{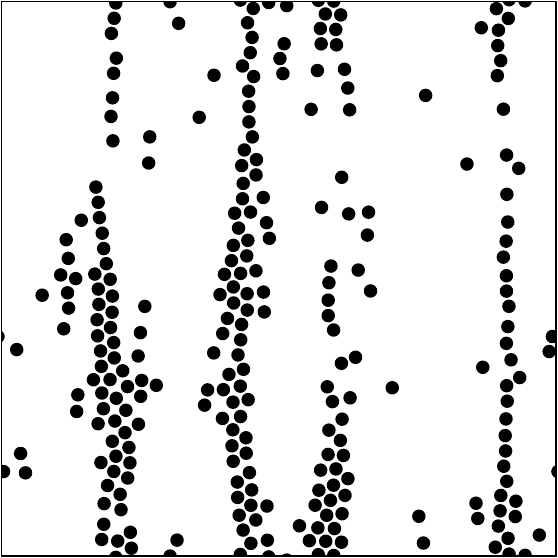}
        \includegraphics[width=0.12\linewidth]{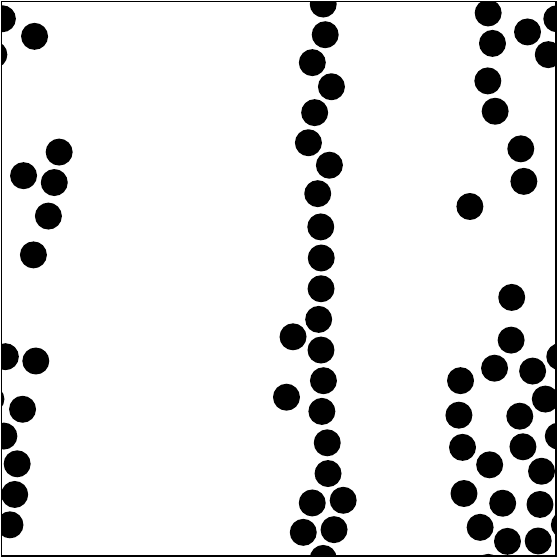} 
        \includegraphics[width=0.09\linewidth]{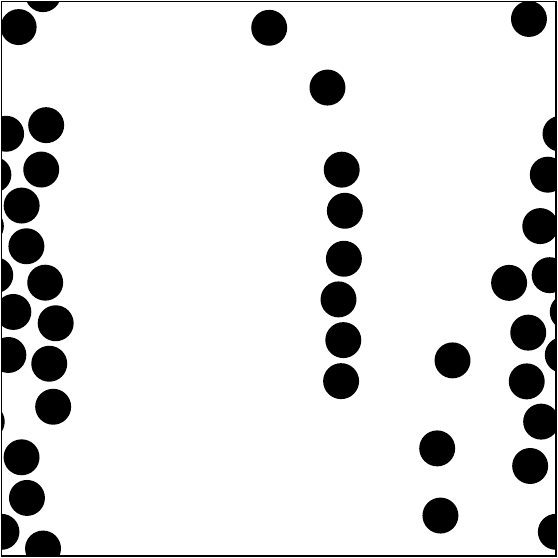} 
        \includegraphics[width=0.06\linewidth]{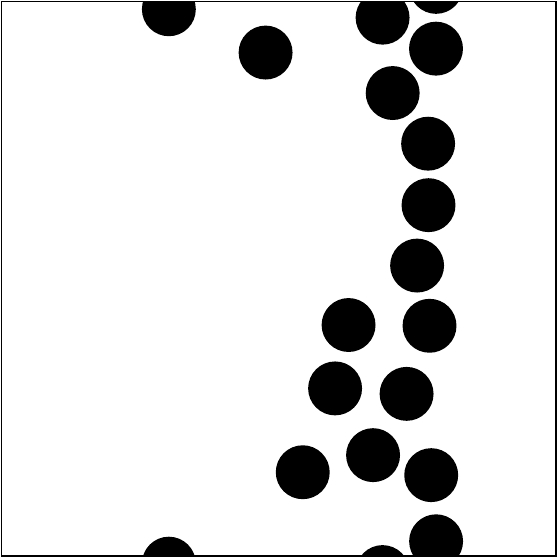} 
        \includegraphics[width=0.045\linewidth]{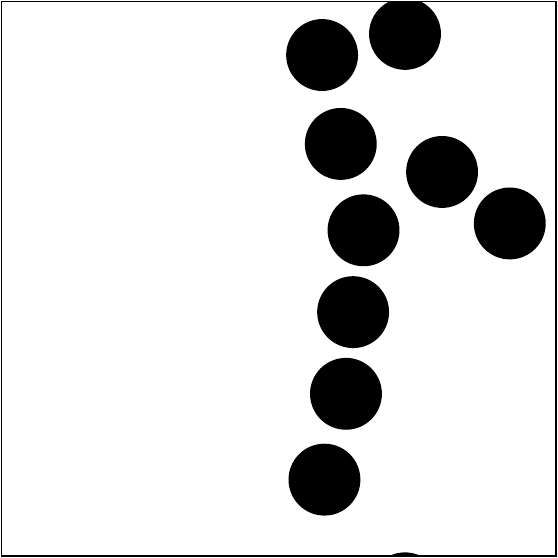} 
        \includegraphics[width=0.03\linewidth]{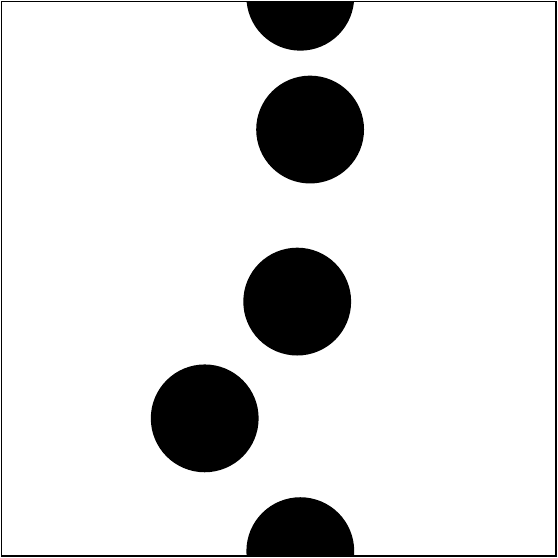} 
    \caption{Generated unit cells with 400, 256, 64, 36, 16, 9, and 4 particles, showing the result with the lowest descriptor loss for each size.}
    \label{fig:ms_images_4_64_400}
\end{figure*}

\begin{figure*}[h]
    \centering

    \begin{subfigure}{\textwidth}
        \centering
        \includegraphics[width=0.22\textwidth]{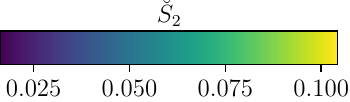}
    \end{subfigure}

    \vspace{2pt}

    \newlength{\imageheight}
    \setlength{\imageheight}{0.4\textwidth}

    \begin{subfigure}{0.4\textwidth}
        \centering
        \begin{minipage}[c][\imageheight][c]{\linewidth}
            \centering
            \includegraphics[width=\linewidth]{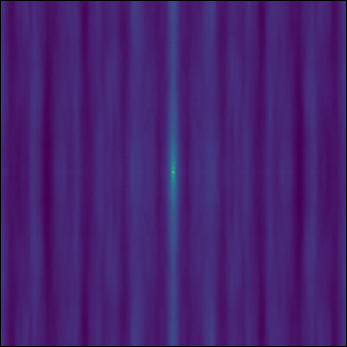}
        \end{minipage}
        \subcaption{TPC of the reference cell}
        \label{fig:optimization_tpc_ref}
    \end{subfigure}
    \quad\quad
    \begin{subfigure}{0.45\textwidth}
        \centering
        \begin{minipage}[c][\imageheight][c]{\linewidth}
            \centering
            \raisebox{-0.5\height}{%
                \includegraphics[width=0.5\linewidth]{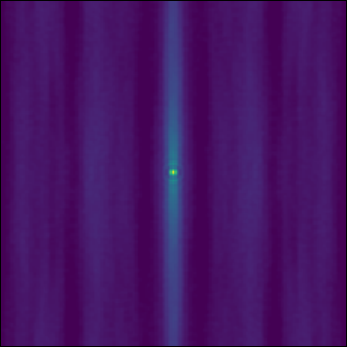}%
            }\hfill%
            \raisebox{-0.5\height}{%
                \includegraphics[width=0.24\linewidth]{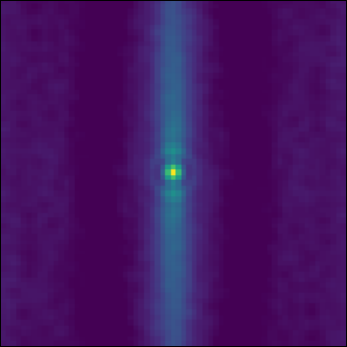}%
            }\hfill%
            \raisebox{-0.5\height}{%
                \includegraphics[width=0.12\linewidth]{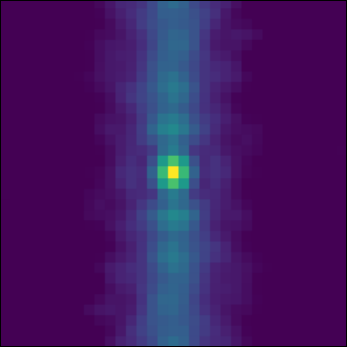}%
            }\hfill%
            \raisebox{-0.5\height}{%
                \includegraphics[width=0.06\linewidth]{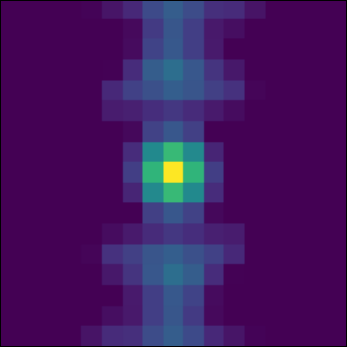}%
            }%
        \end{minipage}
        \subcaption{TPC of the generated cells with varying size}
        \label{fig:optimization_tpc_generated}
    \end{subfigure}

    \caption{TPC of the periodically extended reference cell with $N^\text{incl}=1716$ and of the generated cells with sizes $N^\text{incl} \in \{400, 64, 16, 4\}$ shown in Fig.~\ref{fig:ms_images_4_64_400}.}
    \label{fig:optimization_tpc}
\end{figure*}

Selected generated cells with 400, 256, 64, 36, 16, 9, and 4  particles are shown in Fig.~\ref{fig:ms_images_4_64_400}.
Most particles form chains that are several particles thick, but some particles remain unaggregated.
The TPCs of four unit cells, each generated with a different number of particles, are compared to the TPC of the reference cell in Fig.~\ref{fig:optimization_tpc}.
The area elements have a TPC similar to that of the reference, although differences are noticeable.

\begin{remark}
        In addition to the optimizer of Malitsky and Mishchen\-ko, we tested a gradient solver with Lipschitz estimate and the Barzilai--Borwein solver~\cite{Barzilai1988}, but none of them consistently reached a lower loss. Furthermore, the approach suggested by Seibert~et~al. \cite{Seibert24}, in which both the descriptor and the minimum distance are included in a single loss function, was tested using the optimizers SLSQP, BFGS, L-BFGS-B, and CG, implemented in SciPy, but no significantly lower loss was reached.
\end{remark}
    
\captionsetup[subfigure]{justification=centering}

\subsection{Homogenization results}
\label{sec:homogenization_results}

To assess the quality of the generated cells, their effective properties are compared to those of the reference cell.
First, we consider only the purely mechanical behavior, specifically the initial elasticity tensor.
Second, we evaluate the purely magnetic behavior in the form of the initial susceptibility tensor.
Both the elasticity and susceptibility tensor are obtained from the homogenization results of specific load cases.
Finally, we compare the effective mechanical stresses and magnetizations resulting from a magneto-mechanically coupled load case.
All FE simulations are performed with the Python package NGSolve \cite{ngsolve_62402}.
To investigate the effective behavior of a two-dimensional cell 
$\Omega^{\text{cell}} \subset \mathbb{R}^2$, we restrict the primary 
fields to a plane-strain kinematic and in-plane magnetic state, with 
all field variables independent of $X_3$, and prescribe only macroscopic 
deformations and magnetic inductions in the $X_1$-$X_2$ plane.

\subsubsection{Elasticity tensor}
\label{sec:Stiffness_tensor}

To evaluate the mechanical behavior of the generated two-dimensional cells, we determine the stresses resulting from three linearly independent, purely mechanical (i.e., $\bar{\ve{B}} = \ve{0}$) load cases.
Since we are interested in the initial elasticity tensor, we assume small strains and linear elasticity, i.e., $\bar{\boldsymbol{\upsigma}} = \bar{\tte{C}} : \bar{\boldsymbol{\upvarepsilon}}$ with the elasticity tensor $\bar{\tte{C}}$.
The first two load cases
\begin{equation}
	\bar{\te{F}}^\text{lc1} = \te{1} + \bar{\gamma} \ve{e}_1 \otimes \ve{e}_1
	\quad\text{and}\quad
	\bar{\te{F}}^\text{lc2} = \te{1} + \bar{\gamma} \ve{e}_2 \otimes \ve{e}_2
\end{equation}
lead, for $\bar{\gamma} \ll 1$, to uniaxial strain in the $X_1$- and $X_2$-directions, respectively, i.e., to the linearized strain tensors
\begin{equation}
	\bar{\boldsymbol{\upvarepsilon}}^\text{lc1} = \bar{\gamma} \ve{e}_1 \otimes \ve{e}_1
	\quad\text{and}\quad
	\bar{\boldsymbol{\upvarepsilon}}^\text{lc2} = \bar{\gamma} \ve{e}_2 \otimes \ve{e}_2 .
\end{equation}
The third load case
\begin{equation}
    \label{eq:F_loadcase_3}
	[\bar{F}_{\alpha\beta}^\text{lc3}] = 
	\begin{bmatrix}
		\frac{\bar{\gamma}}{2\bar{\delta}} & \bar{\delta} & 0 \\
		\bar{\delta} & \frac{\bar{\gamma}}{2\bar{\delta}} & 0 \\
		0 & 0 & 1
	\end{bmatrix}
	\quad\text{with}\quad
	\bar{\delta} = \frac{\bar{\gamma}}{\sqrt{2\left(1+\sqrt{1-\bar{\gamma}^2}\right)}}
\end{equation}
leads, for $\bar{\gamma} \ll 1$, to the pure shear linearized strain tensor
\begin{equation}
	[\bar{\varepsilon}_{\alpha\beta}^\text{lc3}] = 
	\begin{bmatrix}
		0 & \frac{\bar{\gamma}}{2} & 0 \\
		\frac{\bar{\gamma}}{2} & 0 & 0 \\
		0 & 0 & 0
	\end{bmatrix}.
\end{equation}
\begin{remark}
    The expression for $\bar{\delta}$ in Eq.~\eqref{eq:F_loadcase_3} is analytically identical to $\bar{\delta} = \sqrt{\frac{1}{2}\left(1 - \sqrt{1 - \bar{\gamma}^2}\right)}$ from Kalina~\cite{Kalina21Diss} for $\bar{\gamma} > 0$, but numerically more stable for small values of $\bar{\gamma}$.
\end{remark}
In the following, the value of $\bar{\gamma}$ is set to $\bar{\gamma} = 10^{-5}$.
The corresponding homogenized stresses according to Sec.~\ref{sec:Finite_strain_magneto-hyperelasticity} are denoted by $\bar{\boldsymbol{\upsigma}}^\text{lc1}$, $\bar{\boldsymbol{\upsigma}}^\text{lc2}$, and $\bar{\boldsymbol{\upsigma}}^\text{lc3}$.
The 3D elasticity tensor in Voigt notation is given by
\begin{equation}
  \begin{aligned}
	\bar{\uuline{C}}^\text{Voigt} &=
	\begin{bmatrix}
		\bar{C}_{1111} & \bar{C}_{1122} & \bar{C}_{1133} & \bar{C}_{1123} & \bar{C}_{1113} & \bar{C}_{1112} \\
		\bar{C}_{2211} & \bar{C}_{2222} & \bar{C}_{2233} & \bar{C}_{2223} & \bar{C}_{2213} & \bar{C}_{2212} \\
		\bar{C}_{3311} & \bar{C}_{3322} & \bar{C}_{3333} & \bar{C}_{3323} & \bar{C}_{3313} & \bar{C}_{3312} \\
		\bar{C}_{2311} & \bar{C}_{2322} & \bar{C}_{2333} & \bar{C}_{2323} & \bar{C}_{2313} & \bar{C}_{2312} \\
		\bar{C}_{1311} & \bar{C}_{1322} & \bar{C}_{1333} & \bar{C}_{1323} & \bar{C}_{1313} & \bar{C}_{1312} \\
		\bar{C}_{1211} & \bar{C}_{1222} & \bar{C}_{1233} & \bar{C}_{1223} & \bar{C}_{1213} & \bar{C}_{1212}
	\end{bmatrix} \\
	&=
	\frac{1}{\bar{\gamma}}
	\begin{bmatrix}
		\bar{\sigma}_{11}^{\mathrm{lc1}} &
		\bar{\sigma}_{22}^{\mathrm{lc1}} &
		\bar{\sigma}_{33}^{\mathrm{lc1}} &
		0 &
		0 &
		\bar{\sigma}_{12}^{\mathrm{lc1}}
		\\[2pt]
		\bar{\sigma}_{22}^{\mathrm{lc1}} &
		\bar{\sigma}_{22}^{\mathrm{lc2}} &
		\bar{\sigma}_{33}^{\mathrm{lc2}} &
		0 &
		0 &
		\bar{\sigma}_{12}^{\mathrm{lc2}}
		\\[2pt]
		\bar{\sigma}_{33}^{\mathrm{lc1}} &
		\bar{\sigma}_{33}^{\mathrm{lc2}} &
		\circ &
		\circ &
		\circ &
		\bar{\sigma}_{33}^{\mathrm{lc3}}
		\\[2pt]
		0 & 0 & \circ & \circ & \circ & 0
		\\[2pt]
		0 & 0 & \circ & \circ & \circ & 0
		\\[2pt]
		\bar{\sigma}_{12}^{\mathrm{lc1}} &
		\bar{\sigma}_{12}^{\mathrm{lc2}} &
		\bar{\sigma}_{33}^{\mathrm{lc3}} &
		0 &
		0 &
		\bar{\sigma}_{12}^{\mathrm{lc3}}
	\end{bmatrix}
	\label{eq:C_voigt}
  \end{aligned}
\end{equation}
after setting some components to $0$ to fulfill $\bar{\sigma}_{13} = \bar{\sigma}_{23} = 0$.
The components marked by $\circ$ are unknown, but have no influence on the resulting stress for a given plane strain state.

To visualize the mechanical anisotropy, the projection of the elasticity tensor in different directions defined by the angle $\phi$ is chosen.
The uniaxial strain $\bar{\varepsilon} = \bar{\gamma} \ve{n} \otimes \ve{n}$ is applied in the direction $\ve{n}(\phi) = \cos{\phi} \, \ve{e}_1 + \sin{\phi} \, \ve{e}_2$.
Note that this prescribed load state differs from the uniaxial stress state used to derive the elastic surface \cite{Böhlke2001}.
By calculating the projection of the deviatoric stress $\bar{\sigma}_0^\text{dev} = \ve{n} \cdot \op{dev}\bar{\boldsymbol{\upsigma}} \cdot \ve{n}$ under the assumption of linear elasticity, i.e., $\bar{\boldsymbol{\upsigma}} = \bar{\tte{C}} : \bar{\boldsymbol{\upvarepsilon}}$, the deviatoric directional stiffness reads
\begin{equation}
	\bar{E}^\text{dev}(\phi) = \frac{\bar{\sigma}_0^\text{dev}}{\bar{\gamma}} = (\ve{n} \otimes \ve{n}) : \op{dev}\left( \bar{\tte{C}} : (\ve{n} \otimes \ve{n}) \right) .
\end{equation}
Likewise, we define the volumetric directional stiffness $\bar{E}^\text{vol}(\phi) = (\ve{n} \otimes \ve{n}) : \op{sph}\left( \bar{\tte{C}} : (\ve{n} \otimes \ve{n}) \right)$
and the total directional stiffness $\bar{E}(\phi) = \bar{E}^\text{dev}(\phi) + \bar{E}^\text{vol}(\phi) = (\ve{n} \otimes \ve{n}) : \bar{\tte{C}} : (\ve{n} \otimes \ve{n})$.

To quantify the relative deviation between two elasticity tensors $\tte{C}$ and $\tte{C}^\text{ref}$, the log-Euclidean distance \cite{Moakher2006}
\begin{equation}
	\epsilon_{\bar{\tte{C}}} = 
	d_\text{log}\left(\bar{\tte{C}}_{2\text{D}}, \bar{\tte{C}}^\text{ref}_{2\text{D}}\right) = 
	\frob{\op{log}\left(\bar{\tte{C}}_{2\text{D}}\right) - \op{log}\left(\bar{\tte{C}}_{2\text{D}}^\text{ref}\right)}
\end{equation}
of the two-dimensional elasticity tensors is computed, which in Voigt notation are given by
\begin{equation}
	\bar{\uuline{C}}^\text{Voigt}_{2\text{D}} = 
	\begin{bmatrix} 
		\bar{C}_{1111} & \bar{C}_{1122} & \bar{C}_{1112} \\ 
		\bar{C}_{2211} & \bar{C}_{2222} & \bar{C}_{2212} \\ 
		\bar{C}_{1211} & \bar{C}_{1222} & \bar{C}_{1212}
	\end{bmatrix} .
\end{equation}
This requires applying the natural logarithm to the tensor's eigenvalues, which are calculated from its Mandel representation.
While the log-Euclidean distance does not directly divide an absolute error by a reference value, it is still a scaling-independent error measure and is expressed as a percentage for convenience.

\begin{figure}[]
    \centering
    \includegraphics[width=1.0\linewidth]{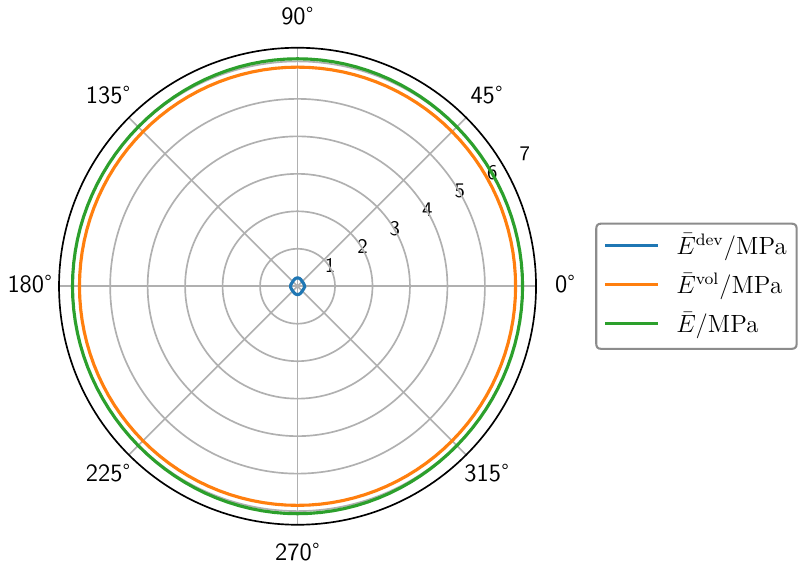}
    \caption{Comparison of the deviatoric, volumetric, and total directional stiffness of the reference cell as a function of the angle $\angle(\ve{n}, \ve{e}_1)$.}
    \label{fig:elasticity_iso_vol}
\end{figure}

In Fig.~\ref{fig:elasticity_iso_vol}, the deviatoric, volumetric, and total directional stiffness of the reference area element shown in Fig.~\ref{fig:particle_segmentation_circles} is depicted as a function of the angle $\angle(\ve{n}, \ve{e}_1)$.
Due to the near-incompressibility of the matrix, the total directional stiffness is much larger than the deviatoric directional stiffness and appears almost isotropic.
In contrast, the deviatoric directional stiffness exhibits a high degree of anisotropy and is therefore used for the stiffness visualizations of the generated cell in Fig.~\ref{fig:results_stiffness}.
We observe that the highest stiffness occurs in the direction of the particle chains.
As the stiffness direction $\ve{n}$ deviates from the chain direction, the stiffness first decreases, but then increases around the perpendicular direction of the particle chains.
This behavior is also observed in the elastic surface of an idealized three-dimensional chain structure \cite{Roth2026}.
In Tab.~\ref{tab:errors_homogenization}, the stiffness error generally decreases as the number of particles in the generated cells increase.

\subsubsection{Magnetic susceptibility tensor}
\label{sec:Magnetic_susceptibility_tensor}

\begin{figure*}[]
    \centering

    \begin{subfigure}{0.4\textwidth}
        \centering
        \includegraphics[width=\linewidth]{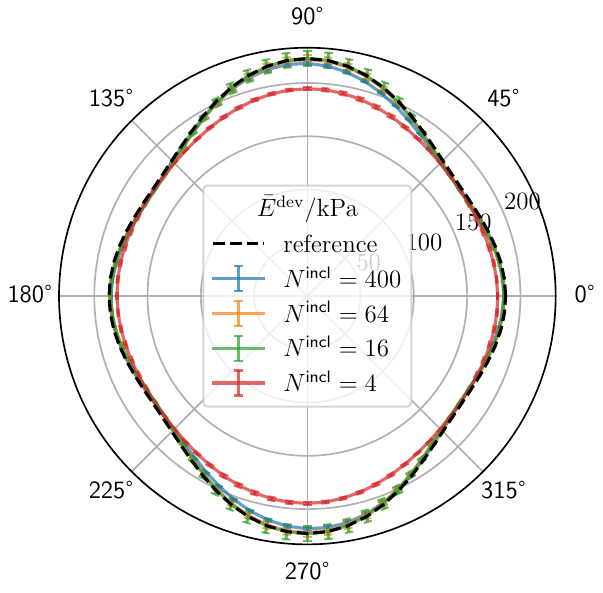}
        \caption{Directional deviatoric stiffness}
        \label{fig:results_stiffness}
    \end{subfigure}
    \hfill
    \begin{subfigure}{0.4\textwidth}
        \centering
        \includegraphics[width=\linewidth]{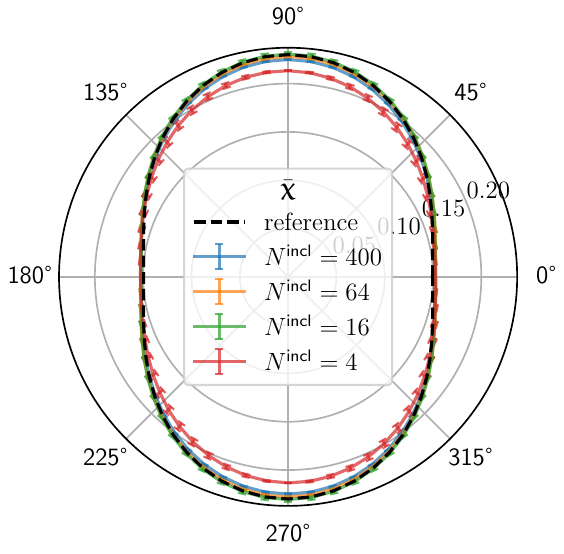}
        \caption{Directional susceptibility}
        \label{fig:results_susceptibility}
    \end{subfigure}

    \caption{(a) Directional deviatoric stiffness and (b) susceptibility as a function of the angle $\angle(\ve{n}, \ve{e}_1)$ for generated unit cells with different numbers of inclusions $N^\text{incl}$. The solid lines denote the mean and the error bars the standard deviation over the five generated cells with the lowest loss.}
    \label{fig:results_stiffness_susceptibility}
\end{figure*}

The magnetic anisotropy is characterized by the effective susceptibility tensor $\bar{\te{\upchi}}$, representing the initial magnetic properties.
To compute this quantity, the two purely magnetic (i.e., $\te{F} = \te{1}$) load cases
\begin{equation}
	\bar{\ve{B}}^\text{lc4} = \bar{\beta} \ve{e}_1
	\quad\text{and}\quad
	\bar{\ve{B}}^\text{lc5} = \bar{\beta} \ve{e}_2
\end{equation}
need to be evaluated.
In the following, a perturbation of $\bar{\beta} = \SI{1e-3}{\tesla}$ is chosen.
The corresponding homogenized magnetic field strengths, computed using the FEM, are denoted by $\bar{\ve{H}}^\text{lc4}$ and $\bar{\ve{H}}^\text{lc5}$.
The 2D inverse permeability tensor
\begin{equation}
	[\bar{\te{\upmu}}^{-1}] =
	\frac{1}{\bar{\beta}}
	\begin{bmatrix}
		\bar{H}_{1}^\text{lc4} & \bar{H}_{2}^\text{lc4} \\[2pt]
		\bar{H}_{2}^\text{lc4} & \bar{H}_{2}^\text{lc5}
	\end{bmatrix}
\end{equation}
allows the susceptibility tensor
\begin{equation}
	\bar{\te{\upchi}} = \te{1} - \mu_0 \bar{\te{\upmu}}^{-1}
\end{equation}
and its directional projection
\begin{equation}
	\bar{\upchi} = \ve{n} \cdot \bar{\te{\upchi}} \cdot \ve{n}
\end{equation}
to be calculated.
The susceptibility tensor characterizes the magnetic relation $\bar{\ve{m}} = \frac{1}{\mu_0}\bar{\te{\upchi}} \cdot \bar{\ve{b}}$ in the linear regime.
The log-Euclidean distance
\begin{equation}
	\epsilon_{\bar{\te{\upchi}}} = 
	d_\text{log}\left(\bar{\te{\upchi}}, \bar{\te{\upchi}}^\text{ref}\right) = 
	\frob{\op{log}\left(\bar{\te{\upchi}}\right) - \op{log}\left(\bar{\te{\upchi}}^\text{ref}\right)}
\end{equation}
is used as a relative error measure in Tab.~\ref{tab:errors_homogenization}.
There, it is shown that the susceptibility error averaged over the five generated cells with the lowest loss decreases with increasing particle count until 16 particles are reached and then remains relatively constant.
The directional susceptibilities of the reference and the generated cells are shown in Fig.~\ref{fig:results_susceptibility}.
It can be seen that the susceptibility is highest in the direction of the particle chains and decreases with increasing deviation from this direction.

\subsubsection{Magneto-mechanically coupled behavior}
\label{sec:Magneto-mechanically_coupled_behavior}

For the evaluation of the magneto-mechanical behavior, we apply a load case with no macroscopic deformation, i.e., $\bar{\te{F}} = \te{1}$, and an effective magnetic induction up to $\bar{\ve{B}} = \frac{\sqrt{2}}{2}(\ve{e}_1 + \ve{e}_2)\,\SI{2.5}{\tesla}$ in the diagonal direction.
The errors of the mechanical stress and the magnetization are given by
$\epsilon_{\bar{\boldsymbol{\upsigma}}} = \frac{\frob{\bar{\boldsymbol{\upsigma}} - \bar{\boldsymbol{\upsigma}}^\text{ref}}}{\frob{\bar{\boldsymbol{\upsigma}}^{\text{ref}}}}$
and
$\epsilon_{\bar{\ve{m}}} = \frac{\frob{\bar{\ve{m}} - \bar{\ve{m}}^\text{ref}}}{\frob{\bar{\ve{m}}^{\text{ref}}}}$,
respectively, and are averaged over the five generated cells with the lowest loss in Tab.~\ref{tab:errors_homogenization} for different particle counts.
The stress error decreases with increasing particle count until 16 particles are reached and then stays between \SIrange{2}{3}{\percent}.
A similar trend is observed for the magnetization error, which decreases from above \SI{1}{\percent} to around \SIrange{0.4}{0.7}{\percent}.

\begin{figure*}[h]
    \centering

    \begin{subfigure}{0.48\textwidth}
        \centering
        \includegraphics[height=5cm]{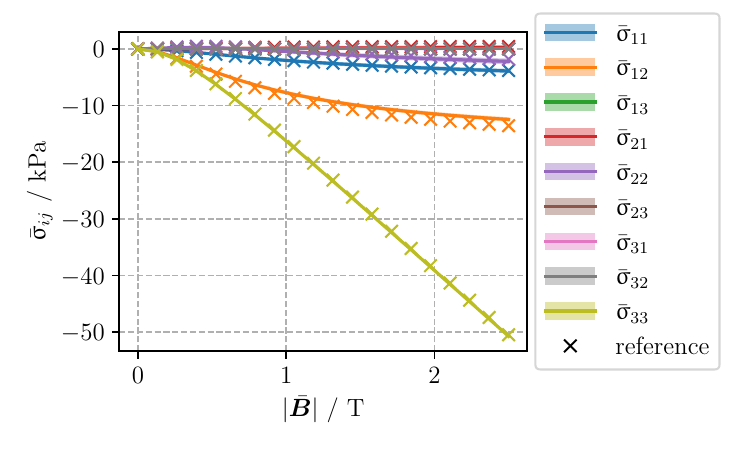}
        \caption{mechanical stress}
        \label{fig:results_sigma}
    \end{subfigure}
    \hfill
    \begin{subfigure}{0.48\textwidth}
        \centering
        \includegraphics[height=5cm]{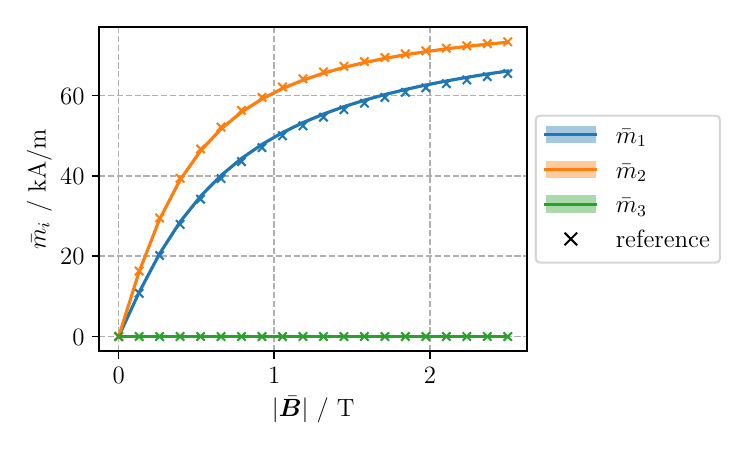}
        \caption{magnetization}
        \label{fig:results_m}
    \end{subfigure}

    \caption{Effective (a) mechanical stress and (b) magnetization for the reference cell and the generated cells with $N^\text{incl}=64$ under the load case $\bar{\te{F}} = \te{1}$ and up to $\bar{\ve{B}} = \frac{\sqrt{2}}{2}(\ve{e}_1 + \ve{e}_2)\,\SI{2.5}{\tesla}$. The solid lines denote the mean and the shaded regions the standard deviation over the five generated cells with the lowest loss.}
    \label{fig:results_sigma_m}
\end{figure*}

\subsubsection{Evaluation of the results}

\begin{table*}[h]
    \begin{center}
        \small
        \caption{Errors of the homogenization results, averaged over the five generated cells with the lowest final loss for each respective size, with respect to those of the reference cell.
        The errors of the mechanical stress tensor $\epsilon_{\bar{\boldsymbol{\upsigma}}}$ and the magnetization $\epsilon_{\bar{\ve{m}}}$ for a magneto-mechanical load case, as well as the errors of the elasticity tensor $\epsilon_{\bar{\tte{C}}}$ and the susceptibility tensor $\epsilon_{\bar{\boldsymbol{\upchi}}}$, are given.
        For comparison, the errors of an isotropic area element with 400 particles placed by the Random Sequential Adsorption algorithm, with identical minimal particle distance and area fraction, are shown.}
        \begin{tabular}{ c | c c c c c c c | c}
        \hline
        & \multicolumn{7}{c|}{\small Generated cell} & \multicolumn{1}{c}{\small Isotropic cell} \\
        $N^\text{incl}$ & 4 & 9 & 16 & 36 & 64 & 256 & 400 & 400 \\
        \hline
        $\epsilon_{\bar{\tte{C}}}$ & 21.29\% & 5.37\% & 5.82\% & 5.09\% & 4.13\% & 4.24\% & 4.14\% & 30.34\% \\
        \hline
        $\epsilon_{\bar{\boldsymbol{\upchi}}}$ & 8.50\% & 4.54\% & 2.10\% & 1.80\% & 2.37\% & 2.64\% & 2.84\% & 30.97\% \\
        \hline
        $\epsilon_{\bar{\boldsymbol{\upsigma}}}$ & 7.13\% & 5.28\% & 2.87\% & 2.07\% & 2.49\% & 2.61\% & 2.78 \%& 26.28\% \\
        \hline
        $\epsilon_{\bar{\ve{m}}}$ & 1.25\% & 1.25\% & 0.51\% & 0.47\% & 0.65\% & 0.68\% & 0.65\% & 5.61\% \\
		\hline
	\end{tabular}
        \label{tab:errors_homogenization}
    \end{center}
\end{table*}

The error measures introduced in Secs.~\ref{sec:Stiffness_tensor} to~\ref{sec:Magneto-mechanically_coupled_behavior} and summarized in Tab.~\ref{tab:errors_homogenization} generally decrease as the particle number increases, though not monotonically.
To assess the error, an isotropic area element with 400 particles is generated using the Random Sequential Adsorption algorithm with the correct area fraction and minimum particle distance, and the errors with respect to the reference microstructure are computed.
For all of the considered effective properties, the best generated cells with at least 64 particles achieve errors approximately one order of magnitude lower than those of the isotropic microstructure.
Using more particles does not systematically decrease the errors.
Therefore, we conclude that cells with 64 particles already accurately capture the effective response of the considered microstructure to the considered loading cases, while reducing the cell area by more than a factor of 6 compared to the non-mirrored original microstructure.
If larger deviations are acceptable in favor of higher computational efficiency, the cell area can be reduced to as few as 16 particles while still roughly representing the reference, resulting in more than 26 times smaller cells.

\section{Summary and outlook}
\label{sec:Summary_and_outlook}

In this work, a procedure to generate two-dimensional unit cells with circular inclusions from a reference microstructure image is presented, along with a corresponding validation strategy.
To this end, the equations for computing the two-point correlation (TPC) of such microstructures are derived, and an optimization problem subject to a minimum inter-particle distance constraint is formulated.
Particles are identified in the reference image, and statistically similar unit cells of varying size are generated. 
The resulting cells are validated against the reference in terms of effective directional stiffness, susceptibility, and response along a magneto-mechanical load path. 
Agreement generally improves with an increasing number of particles: 64 particles yield good agreement, while 16 particles already provide a reasonable approximation.

In future work, the generated cells could serve as an efficient means to produce training data for macroscale material models, which can then be used in macroscale FE simulations \cite{Gebhart2025,Roth2026,Reinken_roadmap_2026}. 
The descriptor optimization itself offers room for improvement, as the loss decreases by less than one order of magnitude despite the large number of iterations. One promising direction is the development of alternative descriptors with more favorable optimization behavior, which could enable faster convergence and lower final losses.
Replacing the circular inclusions by ellipses would allow a closer geometric approximation of the true microstructure. As there is no closed form for the minimum surface distances of ellipses, iterative distance evaluations would be required within the optimization loop.
A principal limitation of this work is the restriction to two-dimensional geometries. An extension to three dimensions is conceptually straightforward once the microstructure is acquired via X-ray microcomputed tomography \cite{Borin2026}, but the computational cost grows substantially and the present optimization scheme may no longer suffice. While the generation of small cells with ellipsoidal inclusions has been demonstrated \cite{Seibert24}, scaling to larger structures remains an open challenge.

\section*{CRediT authorship contribution statement}
\textbf{Heinrich T. Roth:} Conceptualization, Formal analysis, Investigation, Methodology, Visualization, Software, Validation, Writing – original draft, Writing – review \& editing.
\textbf{Lukas Jabs:} Conceptualization, Formal analysis, Methodology, Software, Writing – review \& editing.
\textbf{Matti Schneider:} Resources, Supervision, Funding acquisition, Conceptualization, Formal analysis, Software, Methodology, Writing – review \& editing.
\textbf{Konstantin Gerstenberger:} Investigation, Software, Writing – original draft, Writing – review \& editing.
\textbf{Günter K. Auernhammer:} Resources, Supervision, Funding acquisition, Writing – review \& editing.
\textbf{Karl A. Kalina:} Conceptualization, Methodology, Supervision, Writing – review \& editing.
\textbf{Markus K\"{a}stner:} Resources, Supervision, Funding acquisition, Writing – review \& editing.

\section*{Acknowledgment}
Some of the presented computations were performed on an HPC cluster at the Center for Information Services and High Performance Computing (ZIH) at TU Dresden. 
The authors thus thank the ZIH for generous allocations of computer time.
We thank the Deutsche Forschungsgemeinschaft (German Research Foundation, DFG) for support through the Research Unit FOR 5599 on structured magnetic elastomers, project no. 511114185, via DFG grant reference nos. KA 3309/22-1 and AU 321/16-1, as well as through DFG grant no. 516929769.
The authors would also like to thank Paul Seibert for insightful discussions during the early stages of this work and Jörg Brummund for the helpful discussion on the visualization of elasticity tensors.

\section*{Conflict of interest}
The authors declare no potential conflict of interest.

\section*{Declaration of generative AI and AI-assisted technologies in the manuscript preparation process}
During the preparation of this work, the authors used large language models from OpenAI and Anthropic to improve the readability and language of the manuscript. After using these tools, the authors reviewed and edited the content as needed and take full responsibility for the content of the published article.

\appendix

\section{Microstructure generation with Spectral Density Function}\label{app:Microstructure_generation_with_Spectral_Density_Function}

Alternatively to the TPC, the SDF may be used in the target function for microstructure generation \cite{Bostanabad2018}.
The SDF contains the same information as the TPC, only in Fourier space instead of real space.
This representation might be advantageous for optimization, which is why we investigate it as an alternative descriptor in this section.

To compute the SDF loss, the reference SDF must have the same physical grid size $\mathcal{G}$ as the generated cell.
We reduce the size of the reference SDF by calculating the corresponding TPC and using only the part on $\mathcal{G}$ for the back-transformation, i.e.,
\begin{equation}
	|\hat{\mathcal{I}}_{jk}^\text{ref,red}|^2 =
	\frac{1}{|\mathcal{G}|}
	\sum_{(u,v) \in \mathcal{G}} 
	\left(
		\sum_{(s,t) \in \mathcal{G}^\text{ref}} 
		|\hat{\mathcal{I}}_{st}^\text{ref}|^2 
		\,\mathrm{e}^{\mathrm{i} \boldsymbol{\omega}_{st} \cdot \ve{r}_{uv}}
	\right)
	\,\mathrm{e}^{-\mathrm{i} \boldsymbol{\omega}_{jk} \cdot \ve{r}_{uv}} .
\end{equation}
Note that the zero-frequency component satisfies the equation $|\hat{\mathcal{I}}_{00}|^2 = \phi^2$ on the full grid, whereas in general for $\mathcal{G}^\text{ref} \ne \mathcal{G}$ the relation $|\hat{\mathcal{I}}_{00}^\text{ref,red}|^2 \ne \phi^2$ holds on the reduced grid.
Therefore, the zero frequency of the reference and generated area elements generally will not match. 
Consequently, the loss
\begin{equation}
	\mathcal{L}^\text{SDF}(\underline{\ve{p}}) =
	\sum\limits_{(j,k) \in \mathcal{G}_{\setminus 0}}
	\left( |\hat{\mathcal{I}}_{jk}(\underline{\ve{p}})|^2 - |\hat{\mathcal{I}}^\text{ref,red}_{jk}|^2 \right)^2
\end{equation}
is evaluated only on the grid $\mathcal{G}_{\setminus 0} = \mathcal{G} \setminus \{(0,0)\}$ without the zero frequency.
The derivative with respect to the particle center points is given by
\begin{equation}
	\pd{\mathcal{L}^\text{SDF}(\underline{\ve{p}})}{\ve{p}_l} =
	\sum\limits_{(j,k) \in \mathcal{G}_{\setminus 0}} \!\!
	2 \left( |\hat{\mathcal{I}}_{jk}(\underline{\ve{p}})|^2 - |\hat{\mathcal{I}}^\text{ref,red}_{jk}|^2 \right)
	\pd{|\hat{\mathcal{I}}_{jk}|^2}{\ve{p}_l}
\end{equation}
using the derivative of the SDF given in Eq.~\eqref{eq:d_Ijk_d_pl}.
We then generate monodisperse cells analogously to Sec.~\ref{sec:Numerical_examples_Microstructure_generation}.
Averaging over the five generated cells with 64 particles that exhibit the lowest final descriptor loss, we obtain the errors
$\epsilon_{\bar{\tte{C}}} = 5.68\% $,
$\epsilon_{\bar{\boldsymbol{\upchi}}} = 2.04\% $,
$\epsilon_{\bar{\boldsymbol{\upsigma}}} = 2.76\% $, and
$\epsilon_{\bar{\ve{m}}} = 0.60\% $,
which are similar to those obtained for the microstructures generated with the TPC, see Tab.~\ref{tab:errors_homogenization}.
Thus, there appears to be no systematic advantage in using either the SDF or the TPC.

\section{Microstructure with higher area fraction}\label{app:Microstructure_with_higher_area_fraction}

\begin{figure}[]
    \centering
    \begin{subfigure}{0.3940\textwidth}
        \centering
        \fbox{\includegraphics[width=0.99\linewidth]{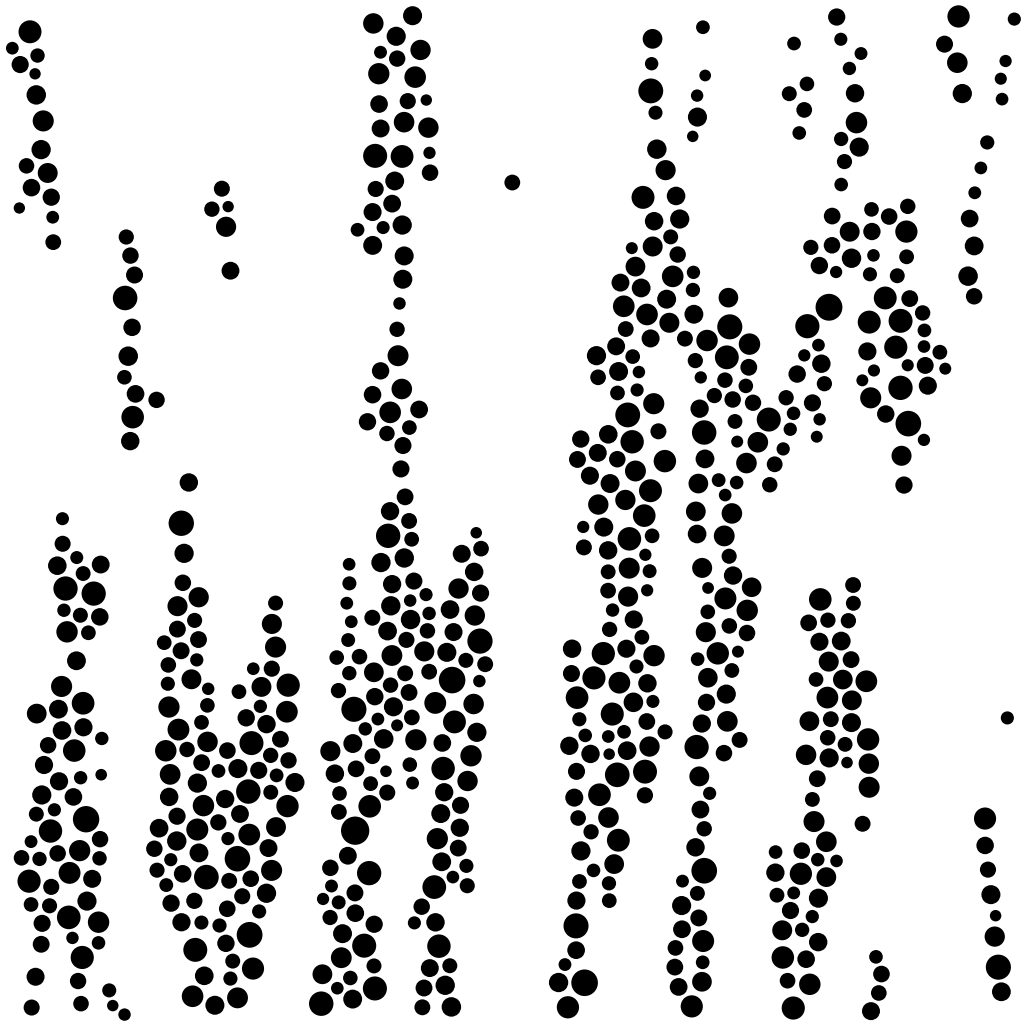}}
        \subcaption{Pre-mirroring reference cell}
		\label{fig:particle_higher_area_fraction_Original}
    \end{subfigure}
	\quad\quad\quad
    \begin{subfigure}{0.4\textwidth}
        \centering
			\includegraphics[width=0.3\textwidth]{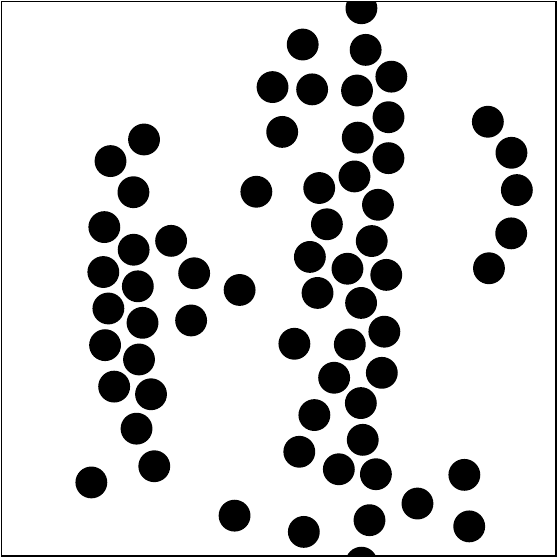}
			\includegraphics[width=0.3\textwidth]{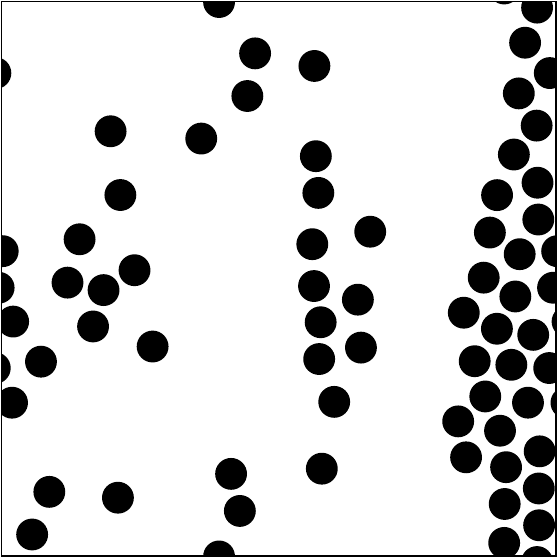}
			\includegraphics[width=0.3\textwidth]{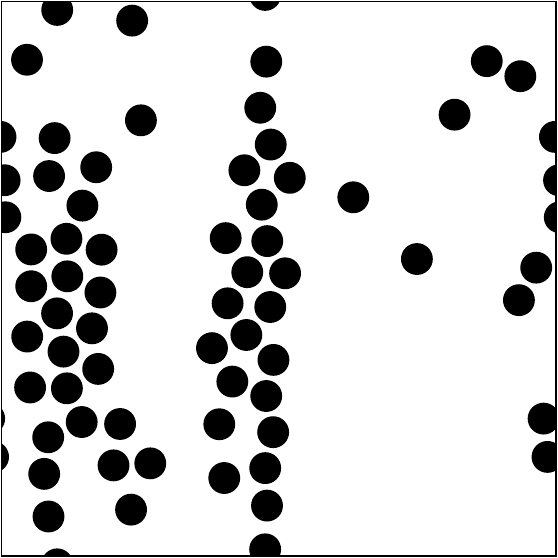}
        \subcaption{Generated cells with 64 particles}
        \label{fig:particle_higher_area_fraction_generated}
    \end{subfigure}
	\caption{Reference microstructure with a higher area fraction before mirroring and the three best generated microstructures with 64 particles each.}
	\label{fig:higher_area_fraction}
\end{figure}

We investigate the applicability of the presented approach to another microstructure with a higher particle area fraction.
Analogously to Sec.~\ref{sec:Determination_of_the_reference_microstructure}, we identify the particles from a microstructure image, which are shown in Fig.~\ref{fig:particle_higher_area_fraction_Original}.
After mirroring the area element to obtain periodicity, we obtain $N^\text{incl} = 2760$ particles.
We choose a minimum particle distance of $d_\text{min} = 0.2 \op{max}(\underline{R}) \approx 2.8$.
The area fraction is $\phi = 0.1657$.
Analogously to Sec.~\ref{sec:Numerical_examples_Microstructure_generation}, we generate 20 monodisperse cells with varying initialization, each with 64 particles.
The three cells with the lowest final loss are shown in Fig.~\ref{fig:particle_higher_area_fraction_generated}.
Homogenization of the five cells with the lowest losses leads to the averaged errors
$\epsilon_{\bar{\tte{C}}} = 7.82\%$,
$\epsilon_{\bar{\boldsymbol{\upchi}}} = 2.09\%$,
$\epsilon_{\bar{\boldsymbol{\upsigma}}} = 2.03\%$, and
$\epsilon_{\bar{\ve{m}}} = 0.45\%$.
The error range is similar to that for the lower area fraction (see Tab.~\ref{tab:errors_homogenization}); only the stiffness error is significantly higher.

\setlength{\fboxrule}{0.5pt}  
\setlength{\fboxsep}{0pt}   

\section{Generation of polydisperse microstructures}\label{app:Generation_of_polydisperse_microstructures}

\begin{figure}[]
	\centering
        \includegraphics[width=0.3\linewidth]{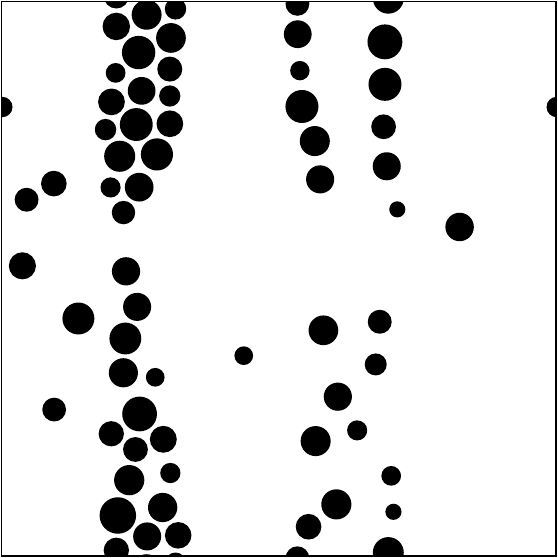} \  
        \includegraphics[width=0.3\linewidth]{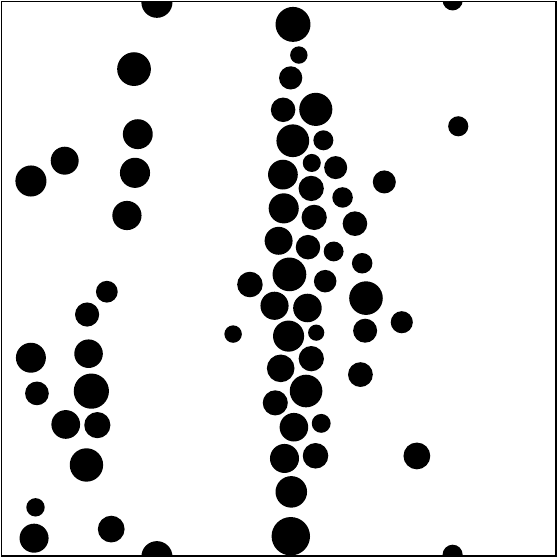} \  
        \includegraphics[width=0.3\linewidth]{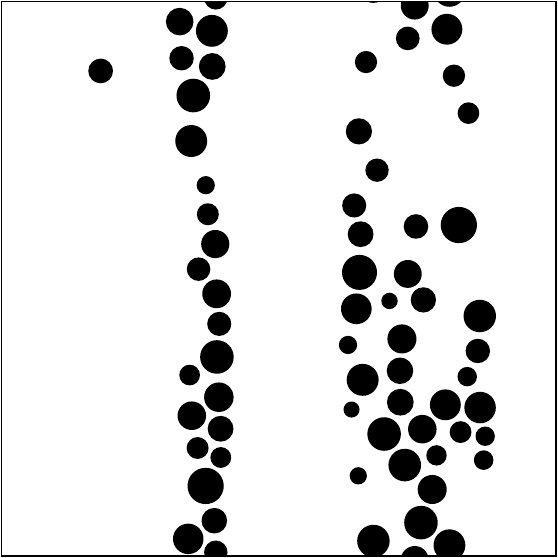} 
    \caption{Generated polydisperse area elements with 64 particles, ordered by loss, with the lowest loss on the left.}
    \label{fig:ms_images_64_polydisperse}
\end{figure}

In the following, we describe the generation of unit cells with 64 polydisperse particles.
Analogously to Sec.~\ref{sec:Numerical_examples_Microstructure_generation}, we generate 20 cells with varying initialization, where the radii are sampled according to Algorithm~\ref{alg:microstructure_generation}, based on the reference microstructure obtained in Sec.~\ref{sec:Particle_segmentation}.
The three cells with the lowest final loss are shown in Fig.~\ref{fig:ms_images_64_polydisperse}.
The lowest loss $\mathcal{L}^\text{TPC}$ is slightly lower than that of the best cell with monodisperse particles, namely $\mathcal{L}^{\text{TPC}} = 7.68 \cdot 10^{-2}$ compared to $7.75 \cdot 10^{-2}$.
However, the errors of the effective properties averaged over the best five cells,
$\epsilon_{\bar{\tte{C}}} = 8.20\%$,
$\epsilon_{\bar{\boldsymbol{\upchi}}} = 2.68\%$,
$\epsilon_{\bar{\boldsymbol{\upsigma}}} = 3.48\%$, and
$\epsilon_{\bar{\ve{m}}} = 0.71\%$,
are worse than those of the monodisperse case; see Tab.~\ref{tab:errors_homogenization}.

\begin{small}
    \bibliographystyle{model1-num-names}
    \bibliography{gen_acc_mic.bib}
\end{small}
\end{document}